\documentclass[review, 12pts]{elsarticle}

\usepackage{amsmath}
\numberwithin{equation}{section}  
\usepackage{graphicx}   
\usepackage{array}      
\usepackage{booktabs}   
\usepackage{verbatim}   
\usepackage{color}      
\usepackage{subfigure}  

\usepackage{pgfplots}

\usepackage{hyperref}
\usepackage{fancyheadings}
\usepackage{mathtools}
\usepackage{amsthm}
\usepackage{amsxtra}
\usepackage{amstext}
\usepackage{amssymb}
\usepackage{enumerate}
\usepackage{longtable}

\usepackage{graphicx}
\usepackage{epstopdf}
\usepackage{epsfig}

\usepackage{setspace}
\usepackage{lineno}

\usepackage{caption}

\begin{document}


\begin{frontmatter}

\title{A realistic transport model with pressure dependent parameters for gas flow in tight porous media with application to determining shale rock properties}

\author{Iftikhar Ali and Nadeem A. Malik\fnref{myfootnote}}
\address{Department of Mathematics and Statistics, King Fahd University of Petroleum and Minerals,
P.O. Box 5046, Dhahran 31261, Saudi Arabia.}
\fntext[myfootnote]{Corresponding: namalik@kfupm.edu.sa  and nadeem$\_$malik@cantab.net}


\begin{abstract}
Shale gas recovery has seen a major boom in recent years due to the
increasing global energy demands; but the extraction technologies are
very expensive. It is therefore important to develop realistic transport
modelling and simulation methods, for porous rocks and porous media, 
that can compliment the field work. Here, a new nonlinear transport model 
for single phase gas flow in tight porous media is derived, incorporating 
many important physical processes that occur in such porous systems: 
continuous flow, transition flow, slip flow, Knudsen diffusion, adsorption 
and desorption in to and out of the rock material, and a correction for high 
flow rates (turbulence). This produces a nonlinear advection-diffusion type of partial 
differential equation (PDE) with pressure dependent model parameters and 
associated compressibility coefficients, and highly nonlinear apparent convective
flux (velocity) and apparent diffusivity. An important application is
to the determination of shale rock properties, such as porosity and permeability,  by 
history matching of the the simulation results to data from pressure-pulse decay tests 
in a shale rock core sample [Pong K., Ho C., Liu J., Tai Y. Non-linear pressure distribution 
in uniform  microchannels. ASME Fluids Eng. Div. (FED) Vol. 197, 51--56, (1994)]. The 
estimates of the rock porosity and the permeability 
from our model simulations are realistic of shale rocks, more realistic than
obtained from previous  models, and illustrates the  potential of the modelling strategy  
presented here in producing accurate simulations of shale gas flow in tight reservoirs.
\end{abstract}

\begin{keyword}
Shale,  gas, transport, porous media, tight, rocks, pressure, porosity, permeability, 
modelling and simulations.
\end{keyword}

\end{frontmatter}

\setlength\parindent{0pt}
\setlength{\parskip}{10pt}


\section{Introduction: Global Energy Perspective and Shale Gas} \label{gep}

At the current time, global energy demand is met by a number of energy resources 
that includes oil, gas, coal, biomass, nuclear, solar, hydro, wind and other 
renewable energy sources. Hydrocarbon based fuels meet more than $80\%$ of 
the world's energy requirements and the world energy supply will continue to 
depend heavily  on carbon based fuels because they are abundant and inexpensive. 
In order to provide an uninterrupted supply of
hydrocarbon fuels and to meet the future energy demands, new avenues are being
explored, both conventional and unconventional, \citet{wang2014natural}
and \citet{islam2014unconventional}. Among the latter, shale gas recovery has 
attracted a lot of attention in the oil and gas industry recently,
\citet{soeder2012shale} and \citet{arthur2008overview}.

Shales are sedimentary rocks that are found abundantly (40\%) on earth. The 
rock structure in shale reservoirs is complex, possessing micropores, nanpores, 
and also microcracks. Fluid flow in porous media occurs in
interconnected networks of void spaces, see Figure \ref{Fig_01}. In
general, the fluid may be of single phase (either liquid or gas), or two 
phase (both liquid and gas), or even multiphase because solid particles
can also be transported within the pores. Furthermore, the fluid may
contain several chemical species, sometimes deliberately added, such
as methanol, lead, sulfuric acid, silica, water. Solvents are used to
open up channels and pore spaces, and coagulants are added to block them.
New advanced technologies such as hydraulic fracturing and horizontal drilling 
are critical for the extraction of the natural gas,
\citet{estrada2016review}. These are very expensive techniques, so they
are often augmented by transport modelling and simulation methods which
may indicate future flow rates, and may also be used to estimate rock
properties, \citet{ma2014pore}.

Shales have very small pore size compared to conventional rock formations.
Different ranges of pore size are reported in the literature, but typically
lies in the range of $50-200\ nm$, \citet{wang2013compositional},
\citet{nia2013pore}. Gas is stored in the network of pores and a fraction
of the gas is adsorbed into the kerogen material which is the solid organic
material, and a fraction of the gas is trapped inside the fractures which
are the cracks or faults in the rock formation.

Shales also have low porosity, typically in the range $4-15\%$, \citet{vesters2013improved} , 
\citet{darishchev2013simulation}, and extremely low permeability  which makes 
the movement of gas molecules very difficult, \citet{aguilera2010flow, aguilera2013flow}. 
The permeability of shale rocks can vary typically  between $10-2000\  nD$, 
\citet{darishchev2013simulation}. The pressure in shale rock formations typically 
varies in the range $25-60\ MPa$; and the temperature varies
from $325-450\ K$, \citet{wang2014natural} and \citet{wang2016discrete}.

It is not surprising therefore that conventional transport methods based
upon the linear Darcy law in a continuum fails to describe the transport
system adequately. In the first place, several flow regimes have to be
accounted for (slip flow, tranitional flow, surface diffusion, and
Knudsen flow) as well as the adsorption and desorption of the gas from
the rock material. The correlation between the measured permeability and
the apparent permeability that appears in such models is also important and
must be modeled,  \citet{song2016apparent}.

Furthermore, the system is highly pressure dependent. As the gas pressure
depletes, or as the temperature changes inside the reservoirs,  the pore
size itself my change with pressure, and the rate of adsorption and
desorption may also change. Some pores may open up, while others may
close down, implying that the pore network itself may alter and therefore
the permeability and porosity are not constant and must be modeled as a
function of the pressure,  \citet{clarkson2013pore}. Correlations for
these and other physical processes in the systen must be obtained or
estimated, as we will see in Sections \ref{nlgas} and \ref{compcoeff}.

A realistic transport model is an important tool in the petroleum industry
because it allows the future pressure distributions inside reservoirs
to be estimated which assists in policy and planning, and for drilling
strategies, and for estimating future recovery levels, \citet{satter1994integrated}, 
\citet{hanea2015reservoir}, \citet{moyner2015application}. In order to accurately 
model flow of gas in shale rock formations, both the network of induced fractures
and also the porous matrix between the fractures must be considered.
Unfortunately, at the current time, the modelling and simulation methods
are still in their infancy, and the challenges are formidable,  \citet{aziz1979petroleum}, \citet{peaceman2000fundamentals}, \citet{chen2008mathematics, marcondes2010element, marcondes20133d, aybar2014effect, fernandes2014comparison}, not least because there are
a large number of pressure dependent modelling parameters in the system
and the transport equation may be highly nonlinear because the apparent
velocity $U_a$ and apparent diffusivity $D_a$ are nonlinear functions of
the pressure $p({\bf x},t)$,  and of the pressure gradient
$\nabla p({\bf x},t)$.

The aim here is to address the first part of the general transport problem of shale gas flow 
described above, that is we consider single phase gas transport through unfractured rocks. 
We  address the problem of gas transport through fractured rocks \citet{akkutlu2016}
 in future studies. We concentrate on the development of a realistic gas transport
model between such fractures. Some progress has been made recently in
developing such transport models, \citet{cui2009measurements} and 
\citet{civan2011shale};  each model incorporates some of the physical
processes, but falls short of the level of realism that is needed for
current needs. Our aim is to develop a realistic transport model for shale
gas flow which incorporates all the important physical transport processes
in the system, and then to demonstrate its effectiveness by applying it to
determining the rock properties, such as the permeability and the porosity
of shale rock core samples, which is itself a very important part of
research in the petroleum and oil industry, and in the geophysical sciences.

The rest of the paper is organised as follows. In Section \ref{ffpm},
we summaries the important transport processes and model parameters in
tight porous systems. In Section \ref{nlgas}, we derive a new transport
model for gas flow in tight porous media, for three-dimensional and for 
one-dimensional domains. In Section \ref{compcoeff}, we derive expressions
for the various compressibility coefficients associated with all the model
parameters. In Section \ref{numproc}, we describe the numerical procedure.
In Section \ref{resim1}, simulation results are presented using our new
transport model in order to determine rock properties through an inverse
problem of matching simulations for experimental data (also known as history 
matching, see \citet{oliver2011}). We discuss the results and draw conclusions in Section \ref{dac}.

\section{Flow in porous media}\label{ffpm}

\subsection{Modelling transport in porous media}
\label{mtpm}

Fluid flow in porous media is a highly complex phenomenon involving many
variables and many different physical processes, \citet{bird1983definition}, \citet{bear2013dynamics}, \citet{mahdi2015review}, \citet{zhang2015analysis}, \citet{su2015uniform}, \citet{muljadi2015impact},
\citet{ramakrishnan2015measurement}. Unlike conventional fluid flow, such
as flow through pipes, or homogeneous turbulence, where the balance equations (Navier-Stokes) are known and only the properties of the fluid  (viscosity
and density) and the size of the domain and boundary conditions govern the
system, in porous media the balance equations are unknown and the properties of the porous media itself, such as the porosity and the permeability,
also play a leading role, \citet{cui2009measurements}, \citet{chen2015nanoscale}, \citet{civan2011porous}, \citet{geng2016diffusion},  \citet{wang2016discrete}.
Sometimes, you also have to deal with turbulence and with multiphase flow.
It is also possible that the rock properties themselves change in response
to changes in the prevailing conditions such as the pressure and the
temperature; for example, the porosity may change because some pore
passages may become blocked over time, while other passages may opened up.
A general theory for flow in porous media is unknown, so we have to
resort to empirical relationships, like Darcy's law, in order to model
the transport through the porous media, \citet{vafai2005handbook}, \citet{muller2015effect}, \citet{faybishenko2015complex}, \citet{benzerga2015micromechanical}.

Mathematical models that describe the transport of gas through tight
reservoirs are based upon the consideration of the amount of gas that
is transported through the reservoir and the amount of gas that is retained
in it. Such models appear in the form of partial differential equations
(PDE's). The principal parameters upon which most models are based are
the intrinsic rock permeability ($K$) and the rock porosity ($\phi$).
An accurate determination of these properties is therefore essential
for developing transport models for flow through porous media, \citet{freeman2011numerical}, \citet{sun2015gas},
\citet{guo2015study}.

\citet{darcy1856} proposed an empirical linear equation relating the
convective flux (or discharge rate), ${\bf u}$, of the fluid, to the
pressure gradient and the rock properties,
\begin{equation}\label{ch2:darcy}
   {\bf u} = - \frac{K}{\mu} {\bf \nabla} p
\end{equation}
The actual fluid velocity {\bf v} is related to the flux through the
porosity, $\phi$, by ${\bf v}={\bf u}/\phi$.  $p$ is the pressure, $K$
is the rock permeability, $\mu$ the viscosity.

Darcy's law yields good results for laminar flow in high porosity porous
media, but in the case of high velocity flow rate it does not produce
satisfactory results, \citet{prada1999modification}, \citet{xu2015new},
\citet{guo2015experimental}. Several studies have shown that the use of
the mathematical models based on Darcy's law are inadequate to study
transport processes through unconventional porous rocks because different
non-laminar flow regimes occur in tight porous media other than the
continuous (viscous) flow, \citet{thauvin1998network},
\citet{cui2009measurements}, \citet{civan2011shale}, \citet{macini2011laboratory}. For high velocities in porous media, inertial
effects can also become significant. Sometimes a nonlinear inertial term is
added in Darcy's equation, known as the Forchheimer term. This term accounts
for the non-linear behavior of the pressure gradient. \citet{forchheimer1901} introduced it as a quadratic term,
\begin{equation}\label{ch2:for}
   {\bf \nabla}  p = - \frac{\mu}{\kappa} {\bf u}  -   \rho{\bf B}\cdot {\bf u}|{\bf u}|
\end{equation}
Here, $\rho$ denotes the gas density and ${\bf B}$ is a constant tensor of rank two (in the most general
three-dimensional case).
The second term on the right accounts for the non-Darcy effects due to high velocities. Many attempts
have been made to modify Darcy's law through generalized models for Forchheimer's
correction term,  \citet{li2001literature,huang2008applicability}.

\begin{figure}[t]
	\centering
		\includegraphics[width=12cm]{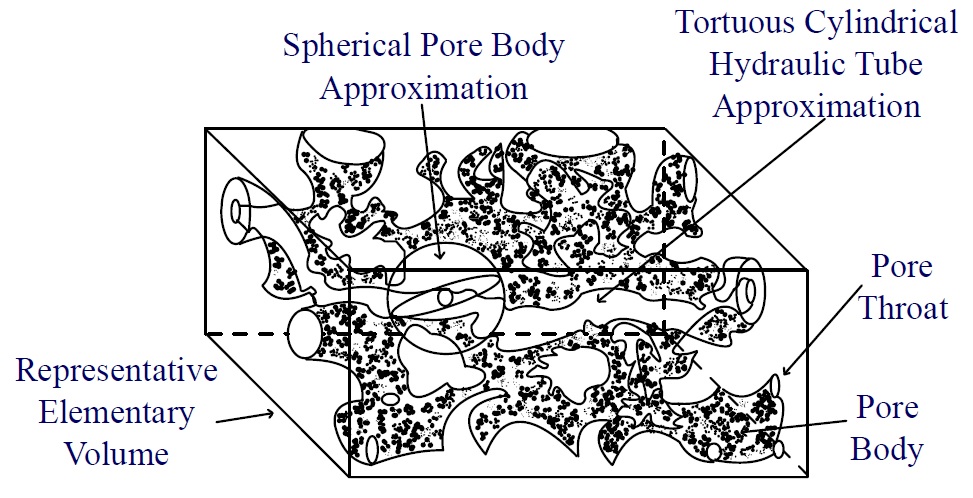}
    \caption{Representative elementary volume  of a porous media which represents the structure
             of the solid matrix. Here the pores are given spherical geometry and flow
             channels are represented as cylindrical tubes.   \citet{civan2001scale}}
	\label{Fig_01}
\end{figure}

\subsection{Classifications of flow regimes based on Knudsen number}
\label{knudsen}

Different flow regimes can be classified through the Knudsen number, \citet{ziarani2012knudsen}, which is defined  as the ratio of the
molecular mean free path $\lambda$ to the hydraulic radius $R_h$, of the
flow channels.
\begin{equation}\label{ch4:KnudsenNumber1}
 K_n = \frac{\lambda}{R_h}.
\end{equation}

The Mean Free Path ($\lambda$) is the average distance traveled by a gas
molecule between collisions with other molecules. There exists several
models for the mean free path, such as given by \citet{loeb2004kinetic}.
\citet{bird1983definition} derived a model, equation  \eqref{ch4:MFP1},
which is based on variable flow head model and showed that it has advantages over the previous studies. \citet{christou2015direct} have recently used
this model in their study of direct simulation Monte Carlo methods in
porous media with varying Knudsen number. The equation for the mean
free path is given by,
\begin{equation}\label{ch4:MFP1}
\lambda = \frac{\mu}{p}\sqrt{\frac{\pi R_g T}{2M_g}},
\end{equation}
where $\rho$ is  gas density (kg/m$^3$), $\mu$ is gas viscosity (Pa-s),
$T$ is the absolute temperature (K),  $R_g=8134$ (J/kmol/K) is the universal gas constant. $p$ is the absolute gas pressure (Pa).

The hydraulic radius $R_h$ is the mean radius of a system of pores and
is given by, \citet{carman1956flow} and  \citet{civan2010effective},
\begin{equation}\label{ch2:rad}
R_h = 2\sqrt{2\tau_h}\sqrt{\frac{K}{\phi}},
\end{equation}
where $\tau_h$ is the tortuosity which is the ratio of apparent length of
the effective mean hydraulic tube to the physical length of the bulk
porous media, and $\phi$ is the porosity which is the fraction of
volume of void spaces to the bulk volume of the porous media, see
Figure  \ref{Fig_01}.

\citet{ziarani2012knudsen}, \citet{rathakrishnan2013gas} and other
researchers have followed  the classification of four flow regimes based
on the Knudsen number. Equations \eqref{ch4:KnudsenNumber1}, \eqref{ch4:MFP1} and \eqref{ch2:rad} yield an expression for the Knudsen number,
\begin{equation}\label{ch2:kn1}
K_n = \frac{\mu}{4p}\sqrt{\frac{\pi R_g T \phi}{M_g \tau_h K }}.
\end{equation}

\begin{figure}[t]
	\centering
		 \includegraphics[width=8cm]{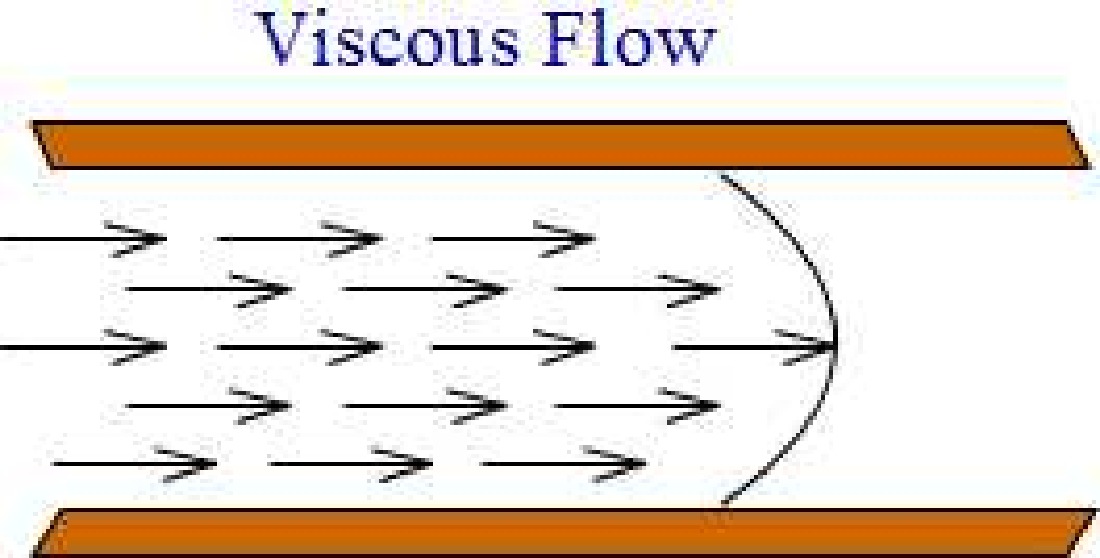}
    \caption{ Viscous flow occurs when the radius of the flow channels is very large
               compared to the mean free path of the gas molecules. Darcy's law is used
               to describe the continuous (laminar) flow.}
	\label{Fig_02}
\end{figure}
\begin{figure}[t]
	\centering
		 \includegraphics[width=8cm]{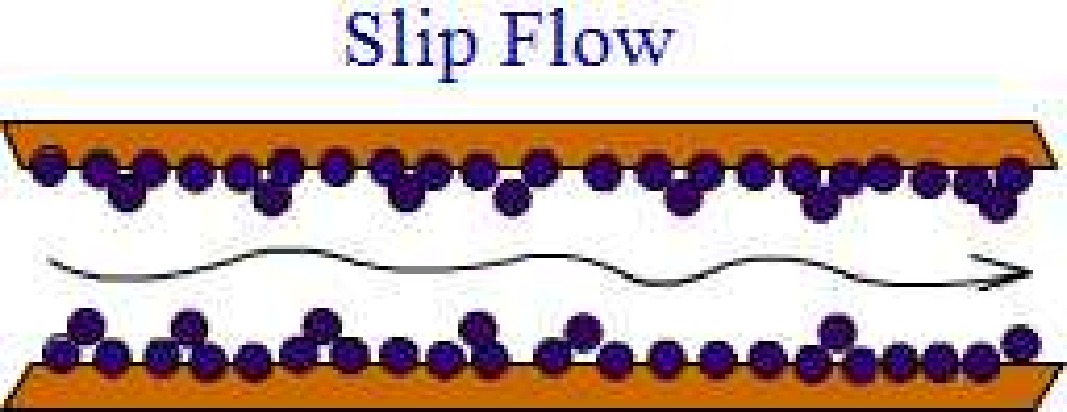}
\caption{Slip flow occurs due to accumulation of gas molecules along pore surface. Transition
            flow occurs When
            more gas molecules collide with the pore surface movement in the gas molecules
            occur because of hopping. Darcy's law starts to fail in the slip and transition
            flow regimes. }
	\label{Fig_03}
\end{figure}
\begin{figure}[t]
	\centering
		 \includegraphics[width=8cm]{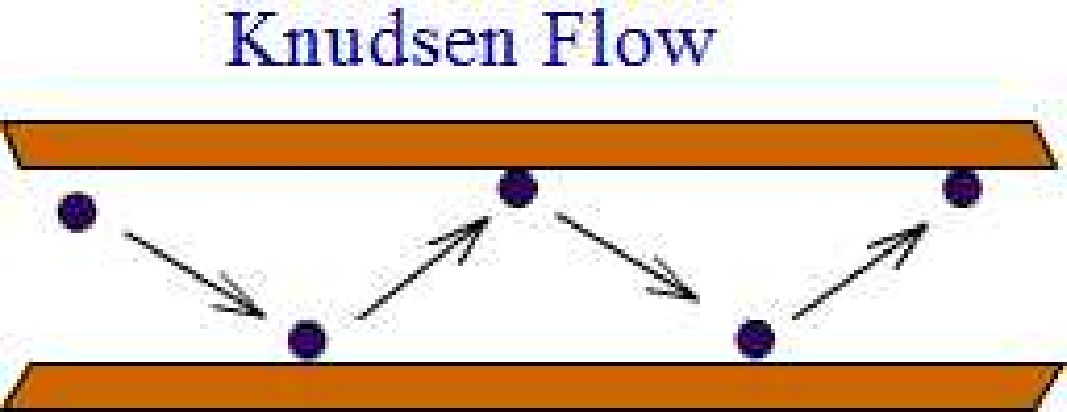}
    \caption{ Knudsen diffusion or free molecular flow occurs when the radius of the flow channels
             is very small compared to the mean free path of the gas molecules. Darcy's law completely
             fails in this regime. }
	\label{fig:Knudsen1}
\end{figure}

The flow regimes are then classified as follows. Continuum (viscous) flow,
\citet{cussler2009diffusion}, exists in the range where $K_n <0.01$, and the conventional Darcy's law can be used to describe the flow.  Darcy's law was derived on the assumption of laminar flow for small Reynolds number $Re \approx O(1)$, Figure \ref{Fig_02}.

Slip flow, \citet{moghaddam2016slip}, exists in the range where
$0.01 < K_n < 0.1$. Gas molecules accumulate along the inside surface of
the pore, Figure \ref{Fig_03}, and they push gas molecules towards
the pore interfaces. Darcy's law, can be employed with some modifications.

Transition flow, \citet{ziarani2012knudsen}, exists in the range where
$0.1 < K_n < 10$. During the slipping phenomenon, when the gas molecules
collide with the gas molecules already stuck to the surface of the porous  
rocks, they exert some force on the molecules and some of the gas molecules
leave the pore surface and become a part of the continuous flow. Conventional 
equations fail and we must use Knudsen diffusion equations.

Knudsen (free molecular) flow, \citet{ziarani2012knudsen}, exists in the range 
where $K_n > 10$. The mean free path of the gas molecules is much
greater than the radius of the flow channels and gas molecules collide
more frequently with the pore walls compared to the collision rate between
gas molecules. It occurs in systems with low pressures or very tight pore
throats as in the case of shale gas or coal bed methane formations, Figure \ref{fig:Knudsen1}.

\subsection{Intrinsic permeability and apparent permeability}
\label{aptperm}

Gas slippage in a porous medium leads to higher than expected measured
gas permeability, the apparent permeability $\mathbf{K}_a$,  compared to
the intrinsic permeability $\mathbf{K}$, \citet{chen2015nanoscale}. Many
correlations between intrinsic and apparent permeabilities have
been proposed in the literature, \citet{klinkenberg1941permeability}, 
\citet{jones1980laboratory}. A formula, derived from Hagen-Poiseuille-type
equation, is given by \citet{beskok1999report}, \begin{equation}\label{ch2:aper} 
{\bf K}_a = {\bf K} f(K_n)
\end{equation}
where $f(K_n)$  is the flow condition function and is given by
\begin{equation}\label{ch2:pcf} 
   f(K_n)=(1+\sigma K_n)\left(1+\frac{4K_n}{1-bK_n}\right),
\end{equation}
where $\sigma$ is called the rarefaction coefficient correlation.
Different correlations for $\sigma$ have been  proposed by 
\citet{beskok1999report, civan2011shale, freeman2011numerical}. In this
work we use the correlation proposed by \citet{civan2010effective},
\begin{equation}\label{ch2:rare}  
\sigma = \sigma_o \left(\frac{K_n^{b_\sigma}}{K_n^{b_\sigma}+a_\sigma}\right)
           = \sigma_o \left({ 1+ \frac{a_\sigma}{K_n^{b_\sigma}} }\right)^{-1}
\end{equation}
where ${a_\sigma}$ and ${b_\sigma}$ are empirical constants, and $b$
in equation \eqref{ch2:pcf} is called the slip factor.


\subsection{Gas adsorption isotherm}
\label{adsorb}

As the gas is transported through the tight pore network, some  of the gas
adheres (clings) to pore surfaces due to the diffusion of gas molecules.
\citet{cui2009measurements} and \citet{civan2011shale} developed a formula
for estimating the amount of adsorbed gas based on Langmuir isotherms,
which will be discussed further in section 4.2.

\section{A transient transport model for gas flow in tight porous media}\label{nlgas}


The approach in developing a realistic transport model is to include as many 
physical processes in the system as possible. Furthermore, a crucial part of 
the new model is to allow all of the model parameters to be
functions of the pressure. For completeness, we first develop a general 
three-dimensional  model, and then to obtain a one-dimensional model from it 
which will be used in subsequent application. The new model encompasses all 
the different flow regimes through the Knudsen number. The turbulent effects
at high velocities are included in the model by a Forchheimer's correction term.

\subsection{Conservation of mass and momentum}
Mass conservation of gas transport through the tight porous media
is described by including the loss of mass of gas by adsorption per
unit bulk volume of porous media and per unit time (the second term
on the left below) and is given by,
\begin{equation}\label{ch3:npm1}
\frac{\partial (\rho \phi)}{\partial t} + \frac{\partial [(1-\phi)q]}{\partial t}
= - \nabla\cdot(\rho \mathbf{u}) + Q
\end{equation}
$q$ is the mass of gas absorbed per solid volume of rock. $Q$ is some external source.

Momentum conservation of gas flowing through porous media is described by
a modified Darcy's law and is  given by \citet{evans1994characterization}
\begin{align}\label{ch3:npm2}
\nonumber  -(\nabla p - \rho g \nabla H) & = \mu \mathbf{K}_a^{-1} \cdot \mathbf{u} +  \rho \mathbf{B} |\mathbf{u}| \cdot \mathbf{u}\\
      & = \mu \mathbf{K}_a^{-1} \left( \mathbf{I}  + \frac{\rho}{\mu}  \mathbf{B}  \mathbf{K}_a |\mathbf{u}|  \right) \cdot \mathbf{u}
\end{align}
where $\rho$ (kg/m$^3$) is the density, $\textbf{u}$ (m$^3$/s/m$^2$) is
the volumetric flux, $\mu$ (Pa s) is the dynamic viscosity of the flowing
gas, $g$ (m$^2$/s) is the magnitude of the gravitational acceleration
vector, $H$ (m) is the depth function, $\textbf{K}_a$ (m$^2$) denotes
the apparent permeability tensor of the rock, $p$ is the pressure, and 
$\mathbf{B}$ represents the inertial and turbulence effects where the
velocity is high, see equation \eqref{ch2:for}.

Setting $\mathbf{F}^{-1} =  \left( \mathbf{I}  + \frac{\rho}{\mu} \mathbf{K}_a \mathbf{B}  |\mathbf{u}|  \right) $, 
where $\mathbf{I}$ is the identity matrix, we obtain,
\begin{eqnarray}\label{ch3:npm2b}
   \mathbf{F} &=&  \left( \mathbf{I}  + \frac{\rho}{\mu} \mathbf{K}_a \mathbf{B}  |\mathbf{u}|  \right)^{-1},
\end{eqnarray}
then equation \eqref{ch3:npm2} can be written as
\begin{align}\label{ch3:npm2a}
           -(\nabla p - \rho g \nabla H) & = \mu (\mathbf{F} \mathbf{K}_a)^{-1} \cdot \mathbf{u}
\end{align}
from which it follows that,
\begin{equation}\label{ch3:npm3}
\mathbf{u} = -\frac{1}{\mu} \left( \mathbf{F}\mathbf{K}_a  \right) \cdot (\nabla p - \rho g \nabla H).
\end{equation}
Combining equations \eqref{ch3:npm1} and \eqref{ch3:npm3}, and after rearranging,
we obtain
\begin{align} \label{ch3:npm5}
\nonumber \frac{\partial (\rho \phi)}{\partial t} + \frac{\partial
[(1-\phi)q]}{\partial t}
\nonumber & = \frac{\rho}{\mu}\left( \mathbf{F} \mathbf{K}_a  \right) : \nabla \nabla p + \nabla \cdot \left(
\frac{\rho}{\mu} \left( \mathbf{F} \mathbf{K}_a  \right) \right)\cdot\nabla p \\
 & - \ \nabla\cdot
      \left({\frac{\rho^2 g}{\mu} \left( \mathbf{F} \mathbf{K}_a  \right) \cdot  \nabla H}\right)  + Q
\end{align}


\subsection{The pressure equation}
When pressure is applied, it changes the physical properties of the system. The 
changes in the physical quantities is measured in terms of compressibility coefficients. 
The compressibility of some parameter, $\gamma(p)$, is the relative change in $\gamma(p)$
in response to the change in the pressure. The isothermal coefficient of
compressibility, $\zeta_{\gamma}(p)$, of property $\gamma$ is defined as,
\begin{equation} \label{ch3:npm7}
\zeta_{\gamma}(p) = \frac{1}{\gamma} \frac{\partial \gamma}{\partial p}
=  \frac{\partial }{\partial p} (\ln\gamma ).
\end{equation}
Thus the isothermal coefficient of compressibility for fluid density $\rho$ is,
\begin{equation} \label{ch3:npm8}
\zeta_{\rho}(p) = \frac{1}{\rho}\frac{\partial \rho}{\partial
p} = \frac{1}{p} - \frac{1}{Z}\frac{\partial Z}{\partial p} = \frac{1}{p} - \zeta_{Z}(p); \quad
{\rm where} \quad \rho=\frac{p M_g }{ZR_g T}.
\end{equation}
For quantities, such as  the fluid viscosity ($\mu$), and the rock porosity ($\phi$),
we assume an exponential integral relation. Thus
\begin{equation} \label{ch3:npm9}
\zeta_{\mu}(p)=\frac{1}{\mu}\frac{\partial \mu}{\partial p}; \quad
\ \text{ where } \ \mu = \mu_{0}\exp\left(\int_{p_0}^p\zeta_{\mu}(p)dp\right)
\end{equation}
and,
\begin{equation} \label{ch3:npm10}
\zeta_{\phi}(p) = \frac{1}{\phi}\frac{\partial \phi}{\partial p};
\quad \ \text{ where } \ \phi = \phi_{0}\exp\left(\int_{p_0}^p\zeta_{\phi}(p)dp\right)
\end{equation}
For a matrix (tensor) quantity like the rock permeability, $\mathbf{K}_a$ ,
we define $\zeta_{\mathbf{K}_a}$ as follows,
\begin{equation} \label{ch3:npm11}
\zeta_{\mathbf{K}_a}(p) \mathbf{I} = \mathbf{K}_a^{-1} \frac{\partial}{\partial p}\left(\mathbf{K}_a \right); \quad  \ \text{ where } \
\mathbf{K}_a = (\mathbf{K}_a)_{0}\exp\left(\int_{p_0}^p\zeta_{\mathbf{K}_a}(p)dp\right)
\end{equation}

Other compressibility coefficients are derived as follows,
\begin{equation} \label{ch3:npm12}
   \zeta_{1}(p) = \frac{1}{\rho\phi}\frac{\partial (\rho\phi)}{\partial p} = \frac{1}{\rho}\frac{\partial \rho}{\partial
   p} + \frac{1}{\phi}\frac{\partial \phi}{\partial p} = \zeta_{\rho}(p) + \zeta_{\phi}(p)
\end{equation}

\begin{equation} \label{ch3:npm13}
\zeta_{2}(p) = \frac{1}{(1-\phi)q}\frac{\partial
[(1-\phi)q}{\partial p} = \zeta_q(p) - \left(\frac{\phi}{1-\phi}\right)\zeta_{\phi}(p)
\end{equation}

\begin{align} \label{ch3:npm14}
\nonumber \zeta_{3}(p) \mathbf{I} & = \left( \frac{\rho}{\mu} \left( \mathbf{F} \mathbf{K}_a \right) \right)^{-1}
\frac{\partial}{\partial p} \left(\frac{\rho}{\mu}\left( \mathbf{F} \mathbf{K}_a \right) \right)  \\
\nonumber & = [\zeta_{\rho}(p) - \zeta_{\mu}(p)   + \zeta_{\left( \mathbf{F} \mathbf{K}_a \right)} (p) ] \mathbf{I} \\
\Rightarrow  \zeta_{3}(p) & = \zeta_{\rho}(p) - \zeta_{\mu}(p)   + \zeta_{\left( \mathbf{F} \mathbf{K}_a \right)} (p),
\end{align}
where,
\begin{align} \label{ch3:npm14a}
\nonumber \zeta_{\left( \mathbf{F} \mathbf{K}_a \right)}(p) \mathbf{I} & = \left( \mathbf{F} \mathbf{K}_a \right)^{-1}
                 \frac{\partial}{\partial p}\left( \mathbf{F} \mathbf{K}_a \right) \\
\nonumber & = [ \mathbf{K}_a^{-1} \zeta_{\mathbf{F}}(p) \mathbf{K}_a + \zeta_{\mathbf{K}_a}(p)] \mathbf{I} \\
\Rightarrow \zeta_{\left( \mathbf{F} \mathbf{K}_a \right)}(p) & = \zeta_{\mathbf{F}}(p) + \zeta_{\mathbf{K}_a}(p).
\end{align}
Equation \eqref{ch3:npm14} then becomes
\begin{align}\label{ch3:npm14b}
\zeta_{3}(p) = \zeta_{\rho}(p) - \zeta_{\mu}(p)   + \zeta_{\mathbf{F}}(p) + \zeta_{\mathbf{K}_a} (p).
\end{align}

From equation \eqref{ch3:npm2b},  we have
\begin{eqnarray}\label{ch3:npm15a}
  \mathbf{F}^{-1} - \mathbf{I} &=&      \frac{\rho}{\mu} \mathbf{K}_a \mathbf{B}  |\mathbf{u}|
\end{eqnarray}
\begin{align}\label{ch3:npm15}
   \zeta_{\mathbf{F}}(p) \mathbf{I} &=  \mathbf{F}^{-1} \frac{\partial } {\partial p} \mathbf{F}
\end{align}
Using the general relation, $\frac{\partial } {\partial p} \mathbf{A}^{-1} = - \mathbf{A}^{-1} \frac{\partial \mathbf{A}} {\partial p} \mathbf{A}^{-1}$, for some tensor $\mathbf{A}$, this leads to,
\begin{align}\label{ch3:npm15b}
\nonumber  \zeta_{\mathbf{F}}(p) \mathbf{I} & = - \mathbf{F}^{-1}  \mathbf{F} \left( \frac{\partial } {\partial p}\mathbf{F}^{-1} \right)\mathbf{F} \\
          &=  -  \frac{\rho}{\mu} \left( \mathbf{K}_a \mathbf{B}  \right)  |\mathbf{u}| \left[ \left(\frac{\rho}{\mu}\right)^{-1}
                 \frac{\partial } {\partial p} \left(\frac{\rho}{\mu} \right) \mathbf{I} + \left( \mathbf{K}_a \mathbf{B}  \right)^{-1}
                 \frac{\partial } {\partial p} \left( \mathbf{K}_a \mathbf{B}  \right) + \frac{1}{|\mathbf{u}|}
                 \frac{\partial |\mathbf{u}|} {\partial p} \mathbf{I}       \right] \mathbf{F}
\end{align}
Substituting equation \eqref{ch3:npm15a} in to equation \eqref{ch3:npm15b}, we obtain
\begin{align}\label{ch3:npm17}
   \zeta_{\mathbf{F}}(p) \mathbf{I} &=  (\mathbf{F} - \mathbf{I}) \left[ \zeta_{\rho}(p) - \zeta_{\mu}(p) +
               \zeta_{\left( \mathbf{K}_a \mathbf{B}  \right)}(p)   + \zeta_{|\mathbf{u}|}(p) \right].
\end{align}

The compressibility coefficient $\zeta_{\left( \mathbf{K}_a \mathbf{B}  \right)}(p)$ can be
expressed as the sum of $\zeta_{\mathbf{K}_a}$ and $\zeta_{ \mathbf{B} }$ as follows,
\begin{align}\label{ch3:npm17a}
\nonumber \zeta_{\left( \mathbf{K}_a \mathbf{B}  \right)}(p) \mathbf{I} & = \left( \mathbf{K}_a \mathbf{B} \right)^{-1}
                 \frac{\partial}{\partial p}\left( \mathbf{K}_a \mathbf{B} \right) \\
\nonumber & = [ \mathbf{B}^{-1} \zeta_{\mathbf{K}_a}(p) \mathbf{B} + \zeta_{\mathbf{B}}(p)]  \mathbf{I}\\
\Rightarrow \zeta_{\left( \mathbf{K}_a \mathbf{B}  \right)}(p) & = \zeta_{\mathbf{K}_a}(p) + \zeta_{\mathbf{B}}(p)
\end{align}
where,
\begin{equation}\label{ch3:npm18}
  \zeta_{\mathbf{B}}(p)\mathbf{I} = \mathbf{B}^{-1} \frac{\partial \mathbf{B}}{\partial p}
\end{equation}

Substituting  equation \eqref{ch3:npm17a}  into equation \eqref{ch3:npm17}, we obtain
\begin{equation}\label{ch3:npm18a}
   \zeta_{\mathbf{F}}(p) \mathbf{I} =  (\mathbf{F} - \mathbf{I}) \left[ \zeta_{\rho}(p) - \zeta_{\mu}(p)  +
                              \zeta_{\mathbf{K}_a}(p)  + \zeta_{\mathbf{B}}(p) + \zeta_{|\mathbf{u}|}(p) \right]
\end{equation}

Substituting equation \eqref{ch3:npm18a} into equation \eqref{ch3:npm14a} and then into equation \eqref{ch3:npm14}, we obtain
\begin{align}\label{ch3:npm19}
\nonumber \zeta_{3}(p)\mathbf{I} & = (\zeta_{\rho}(p)  + \zeta_{\mathbf{K}_a}(p) - \zeta_{\mu}(p)) \mathbf{I} + \zeta_{\mathbf{F}}(p)\mathbf{I}  \\
            & = \mathbf{F}[\zeta_{\rho}(p)  + \zeta_{\mathbf{K}_a}(p) - \zeta_{\mu}(p) ] +
             (\mathbf{F} - \mathbf{I}) ( \zeta_{\mathbf{B}}(p)  + \zeta_{\mathbf{|\mathbf{u}|}}(p)  )
\end{align}

Using equation \eqref{ch2:aper}, we have,
\begin{align} \label{ch3:npm20}
\nonumber \zeta_{\mathbf{K}_a}(p) \mathbf{I} & = \mathbf{K}_a^{-1} \frac{\partial }{\partial p}  \mathbf{K}_a
\nonumber                = ( f \mathbf{K} )^{-1}  \frac{\partial  }{\partial p}  f \mathbf{K}  \\
\nonumber                & = [\zeta_{f}(p) + \zeta_{\mathbf{K}} (p) ]  \mathbf{I} \\
\Rightarrow          \zeta_{\mathbf{K}_a}(p) &  = \zeta_{f}(p) + \zeta_{\mathbf{K}} (p)
\end{align}
where,
\begin{equation} \label{ch3:npm21}
\zeta_{\mathbf{K}}(p)\mathbf{I} = \mathbf{K}^{-1} \frac{\partial }{\partial p}  \mathbf{K};
\quad \ \text{ and } \
\mathbf{K} = (\mathbf{K})_0\exp\left(\int_{p_0}^p\zeta_{\mathbf{K}}(p)dp \right),
\end{equation}
and,
\begin{equation} \label{ch3:npm22}
\zeta_{f}(p)  \mathbf{I} = (f(\mathbf{K}_n))^{-1} \frac{\partial }{\partial p} f(\mathbf{K}_n),
\end{equation}
and,
\begin{align} \label{ch3:npm23}
\nonumber \zeta_{|\mathbf{u}|}(p) & = |\mathbf{u}|^{-1} \frac{\partial }{\partial p}  |\mathbf{u}|  \\
                        & = \frac{1}{\mathbf{u} \cdot \mathbf{u}} \left( \mathbf{u} \cdot \frac{\partial \mathbf{u}}{\partial p} \right).
\end{align}

Note that all compressibility coefficients are combinations of four basic ones, namely, $\zeta_\rho,
\zeta_K, \zeta_f,$ and $\zeta_\mu$.

From equations \eqref{ch3:npm12} and \eqref{ch3:npm13}, we derive the following expressions:
\begin{equation} \label{ch3:npm26}
\frac{\partial (\rho\phi)}{\partial t} = \frac{\partial
(\rho\phi)}{\partial p}\frac{\partial p}{\partial t} = \rho\phi
\zeta_{1}(p)\frac{\partial p}{\partial t},
\end{equation}
and
\begin{equation} \label{ch3:npm27}
\frac{\partial (1-\phi)q}{\partial t} = \frac{\partial
(1-\phi)q}{\partial p}\frac{\partial t}{\partial t} = (1-\phi)q
\zeta_{2}(p)\frac{\partial p}{\partial t}.
\end{equation}

Furthermore, we note that,
\begin{align} \label{ch3:npm28}
   \nonumber \nabla \cdot \left( \frac{\rho}{\mu} \left( \mathbf{F} \mathbf{K}_a \right) \right)  &  \equiv
   \left( \nabla p \frac{\partial }{\partial p} \right) \cdot  \left( \frac{\rho}{\mu} \left( \mathbf{F}
   \mathbf{K}_a \right) \right)\\
    & = \zeta_{3}(p)\nabla p \cdot  \left(\frac{\rho}{\mu} \left( \mathbf{F} \mathbf{K}_a  \right) \right)
\end{align}

Substituting equations \eqref{ch3:npm26}, \eqref{ch3:npm27} and \eqref{ch3:npm28} into
\eqref{ch3:npm5}, we obtain
\begin{align} \label{ch3:npm29}
\nonumber \rho\phi \zeta_{1}(p)\frac{\partial p}{\partial t} &+ (1-\phi)q \zeta_{2}(p)\frac{\partial p}{\partial t}   =  \\
\nonumber
  & \frac{\rho}{\mu}\left( \mathbf{F} \mathbf{K}_a  \right) : \nabla \nabla p +   \zeta_3(p)\nabla p \cdot
    \left(\frac{\rho}{\mu} \left( \mathbf{F} \mathbf{K}_a  \right) \right) \cdot \nabla p \\
  & - \nabla\cdot
       \left({\frac{\rho^2 g}{\mu} \left( \mathbf{F} \mathbf{K}_a  \right) \cdot  \nabla H}\right)  + Q
\end{align}

We define the apparent diffusivity $\mathbf{D}_a$ (m$^2$/s) as,
\begin{align} \label{ch3:npm30}
\mathbf {D}_a(p) & = \frac{\rho\mathbf{F} \mathbf{K}_a }{\mu} \chi,
\end{align}
where
\begin{align} \label{ch3:npm30a}
\chi^{-1} & = \left[\phi \zeta_{1}(p) + (1-\phi) \frac{q}{\rho} \zeta_2(p) \right].
\end{align}
Using equations \eqref{ch3:npm28}, \eqref{ch3:npm30} and \eqref{ch3:npm30a}  in equation
\eqref{ch3:npm29} we obtain,
\begin{align}\label{ch3:npm31}
\nonumber \frac{\partial p}{\partial t} & = \mathbf{D}_a(p) : \nabla \nabla p + \zeta_3(p)\nabla p \cdot \mathbf{D}_a(p)  \cdot \nabla p \\
  &  -  \ \rho g \mathbf{D}_a : \nabla \nabla H
      -    \rho g\zeta_3(p) \nabla p \cdot \mathbf{D}_a  \cdot \nabla H
      -    \nabla (\rho g) \cdot \mathbf{D}_a  \cdot \nabla H
     + \chi Q
\end{align}

Equation \eqref{ch3:npm31} is the most general transport equation for gas flow in three-dimensional
porous media which can be derived under the present assumptions, which includes gravity and
a general source term.

The transport model \eqref{ch3:npm31} has a number of important features. It incorporates the
various flow regimes that occur in the porous media, and the high velocity effects are included
through the Forchheimer's nonlinear correction term. The turbulence factor $\mathbf{B}$ is
considered as a function of $\mathbf{K}_a$, $\phi$, and $\tau$.  Moreover, the parameters
$\phi$, $\mu$, $\rho$ are  functions of pressure $p$. Hence the model \eqref{ch3:npm31} has
nonlinear coefficients $D_a$ and $U_a$.


\subsection{No gravity}

Under the assumption of no gravity ($g = 0$), equation  \eqref{ch3:npm31} becomes
\begin{align}\label{ch3:npm32}
\frac{\partial p}{\partial t} & = \mathbf{D}_a(p) : \nabla \nabla p + \nabla p \cdot \mathbf{D}_a(p) \zeta_3(p) \cdot \nabla p + \psi Q.
\end{align}

\subsection{No source term}

Under the further assumption of no source/sink term ($Q = 0$), equation \eqref{ch3:npm32} becomes
\begin{align}\label{ch3:npm32a}
\frac{\partial p}{\partial t} & = \mathbf{D}_a(p) : \nabla \nabla p + \nabla p \cdot \mathbf{D}_a(p) \zeta_3(p) \cdot \nabla p.
\end{align}

\subsection{One-dimensional equation}

Under the further assumption of one-dimensional flow, equation  \eqref{ch3:npm32a} becomes
\begin{equation}\label{ch3:npm33}
\frac{\partial p}{\partial t} + U_a(p,p_x)\frac{\partial p}{\partial x}
= D_a(p) \frac{\partial^2 p}{\partial x^2}
\end{equation}
where,  $U_a$ (m/s) is called the apparent convective flux (or convective velocity) and is defined by,
\begin{equation}\label{ch3:npm34}
U_a = -\zeta_3(p) D_a(p) \frac{\partial p}{\partial x}
\end{equation}
and
\begin{align} \label{ch3:npm34a}
D_a(p) & = \frac{\rho}{\mu} \frac{ F K_a } {(\rho\phi \zeta_{1}(p) + (1-\phi) q \zeta_2(p))} ,
\end{align}
is called the apparent diffusivity, where $F$ and $K_a$ are now scalar quantities.

Similar models were considered by \citet{malkovsky2009new, liang2001nonlinear, civan2011shale}. 
However, these models do not include the high velocity corrections, and some of the models make 
other approximations such as constant model parameters.


\subsection{Non-dimensional steady state pressure equation}

A steady state model can be obtained by setting $\displaystyle{\frac{\partial p}{\partial t}=0}$ in equation \eqref{ch3:npm33} to yield
\begin{equation}\label{ch3:npm35}
U_a(p,p_x)\frac{\partial p}{\partial x} = D_a(p) \frac{\partial^2 p}{\partial x^2},
\end{equation}
which can be rearranged to yield,
\begin{equation}\label{ch3:npm36}
L_a(p,p_x)\frac{\partial p}{\partial x} = \frac{\partial^2 p}{\partial x^2},
\end{equation}
where,
\begin{equation}\label{ch3:npm37}
L_a = - \zeta_3(p) \frac{\partial p}{\partial x},
\end{equation}
where $\zeta_3(p)$ is given by equation \eqref{ch3:npm14b}.

For the case of transport through a core sample of size $x_D$, with pressure given at the inlet and outlet boundaries, a dimensionless form of steady state equation can be obtained by considering the following dimensionless variables,
$$x_D = \frac{x}{L} \qquad \text{and} \qquad p_D(x_D) = \frac{ p(x) - p_d}{p_u - p_d},$$
which yields,
\begin{equation}\label{ch3:npm33a}
L_a(p_D, (p_D)_x) \frac{\partial p_D}{\partial x_D} = \frac{\partial^2 p_D}{\partial x_D^2}, \qquad 0 \leq x_D \leq 1.
\end{equation}

Furthermore, a non-dimensional transient pressure equation can be obtained by introducing the dimensionless variables,
\begin{equation}\label{ch3:npm45}
  x_D = \frac{x}{L},\ \   t_D = \frac{t}{t_0},\ \  p_D(x_D,t_D) = \frac{p(x,t) - p_d(t)}{p_u(0) - p_d(0)}.
\end{equation}
\begin{equation}\label{ch3:npm46}   
  t_0 = \frac{L^2}{D_0}, \ \ U_D = \frac{U_a}{U_0}, \ \ D_D = \frac{D_a}{D_0}, \ \ Pe = \frac{LU_0}{D_0}.
\end{equation}
The transient equation \eqref{ch3:npm33} then reduces to the following dimensionless form,
\begin{equation}\label{ch3:npm47}   
\frac{\partial p_D}{\partial t_D} + Pe U_D \frac{\partial p_D}{\partial x_D}
= D_D \frac{\partial^2 p_D}{\partial x_D^2}, \qquad  0 \leq x_D \leq 1, \qquad t_D > 0
\end{equation}
where $Pe$ is the Peclet number, $D_o$ and $U_o$ are values of the diffusivity
coefficient $D_a$ and the convective flux $U_a$ at some specific pressure, and
$p_u(t)$ and $p_d(t)$ are pressures in the upstream and the downstream reservoirs.


\section{Compressibility coefficients}
\label{compcoeff}

\subsection{Parameters in the new transport model}

In application, we will be dealing with one-dimensional flow, with zero gravity, 
and with no external source, equation \eqref{ch3:npm33}. $U_a$, $D_a$, $\chi$, 
$F$, $\zeta_{1}(p)$, $\zeta_2(p)$, and $\zeta_{3}$ are defined in equations 
\eqref{ch3:npm34}, \eqref{ch3:npm34a}, \eqref{ch3:npm30a}, \eqref{ch3:npm2b},
\eqref{ch3:npm12}, \eqref{ch3:npm13}, and \eqref{ch3:npm19}, respectively. For 
transport through a reservoir, it requires only boundary and initial conditions 
to solve for the pressure distribution $p(x,t)$.

A complication is the appearance of so many compressibility coefficients
each of which must be known or modelled in order for this system to be solvable. 
We list these models below.

The real gas deviation factor ($Z$) is calculated from \citet{mahmoud2014development},
\begin{align}\label{ch4:ZComp1}
Z &= a_Z p_r^2 + b_Z p_r + c_Z, \\
a_Z &= 0.702\exp(-2.5t_r),\\
b_Z &= -5.524\exp(-2.5t_r), \\
c_Z &= 0.044t_r^2 -0.164t_r + 1.15,
\end{align}
where $p_r=p/p_c$ is the reduced pressure and  $t_r=T/T_c$, is the reduced temperature, and
$p_c$ and is critical pressure, and $t_c$, is the critical temperature -- these quantities are assumed
known. The density, $\rho$, of real gases is defined by the relationship, $\rho = \frac{pM_g}{ZR_g t}$,
which can be re-expressed as,
\begin{equation}\label{ch4:gasDensity}
\rho = \frac{p_r M_g}{t_r R_g Z(p_r)}.
\end{equation}

The compressibility coefficient of the gas density, $\zeta_{\rho}$, is thus given by,
\begin{align}\label{ch4:gasComp1}
\zeta_{\rho}(p) & = \frac{d (\ln \rho)}{dp}
                          = \frac{1}{p} - \frac{1}{p_cZ} \left(\frac{2a_Z p}{ p_c} + b_Z \right).
\end{align}
Setting $\zeta_{Z}(p) = \frac{1}{Z}\frac{\partial Z}{\partial p}$, this re-arranges to,
\begin{equation} \label{ch4:gasComp2}
\zeta_{gas}(p) = \zeta_{\rho}(p) +  \zeta_{z}(p) = \frac{1}{p}.
\end{equation}


The model for the gas dynamic viscosity, $\mu$, used here is, \citet{mahmoud2014development},
\begin{align}\label{ch4:gasVisco}
      \mu & = \mu_{S_c} \exp(A_{\mu} \rho^{B_{\mu}} ), \\
  A_\mu & = 3.47+1588t^{-1}+0.0009M_g\\
  B_\mu  & = 1.66378-0.04679A_\mu\\
  \mu_{Sc} & = \frac{1}{10.5^4} \left[{\frac{M_g^3p_c^4}{t_c}}\right]^{1/6}
\end{align}
The compressibility coefficient of gas viscosity, $\zeta_{\mu}$, is then given by
\begin{equation}\label{ch4:CompCoVisco}
\zeta_{\mu}(p) = \frac{d }{dp} \ln\mu  = A_{\mu}B_{\mu} \rho^{B_{\mu}}\zeta_{\rho}(p)
\end{equation}

For the porosity, $\phi$, we use the correlation, \citet{bockstiegel1966porosity}, \citet{walsh1984effect}, \citet{regnet2015influence}, \citet{zheng2015relationships},
\begin{align}\label{ch4:porosity}
\phi & = a_{\phi} \exp(-b_{\phi} p^{c_{\phi}}),
\end{align}
where $a_{\phi}$, $b_{\phi}$, and $c_{\phi}$ are empirical constants, assumed known.
The compressibility coefficient of the porosity, $\zeta_{\phi}$, is given by,
\begin{align}\label{ch4:porosityComp}
\zeta_{\phi}(p) & = \frac{d }{dp} \ln\phi = -b_{\phi}c_{\phi} p^{c_{\phi}-1}
\end{align}


The Intrinsic Permeability, $K$,  is a fundamental property of the reservoir rocks.
Different models and empirical relations
have been proposed  to estimate the permeability of reservoir rocks. One of the most
commonly used relations is the Kozeny-Carman equation which is derived on the assumption
of continuous flow of fluid through a bundle of parallel tubes of constant diameter,
see \citet{xu2008developing}. This equation gives good results for the homogeneous porous
media with the laminar flow, but it fails to accurately predict the permeability
of  heterogeneous reservoirs with a complex network of pores. It also fails for
reservoirs with very low porosity and low permeability values. Here, we use a
modified Power-Law form of the Kozeny-Carman equation, \citet{civan2014improved} ,
\begin{equation}\label{ch4:KozenyCarmanEq1}
  \sqrt{\frac{K}{\phi}} = \Gamma_{KC}\left(\frac{\phi}{\alpha_{KC} - \phi}\right)^{\beta_{KC}},
\end{equation}
where $\alpha_{KC}$, $\beta_{KC}$, and $\Gamma_{KC}$ are empirical constants, with
$\phi < \alpha_{KC} \leq 1$, and $0\leq \beta_{KC} < \infty$, and $\Gamma_{KC} \geq 0$.
This can be rearranged to yield,
\begin{equation}\label{ch4:KozenyCarmanEq2}
  K = \Gamma_{KC}^2 \frac{\phi^{2\beta_{KC} + 1 }}{ (\alpha_{KC} - \phi)^{2\beta_{KC}}}.
\end{equation}
The compressibility coefficient for the intrinsic permeability, $\zeta_{K}$, is given by,
\begin{align}\label{ch4:KCComp}
\zeta_{K}(p) & = \frac{d(\ln K) }{dp} = -b_{\phi}c_{\phi} p^{c_{\phi}-1}\left( 1 + \frac{2\alpha_{KC}\beta_{KC}}{\alpha_{KC} - \phi}\right).
\end{align}

Tortuosity, $\tau$, is a measure of the geometric complexity of the pore network and
inter-connectivity, and it is defined as the ratio of the length of a typical streamline,
or path, between two boundaries, to the the bulk length of the reservoir rock.
There exists several relations between tortuosity and porosity but none of them works for all
situations, see \citet{matyka2008tortuosity}. We use the following correlation for the tortuosity, \citet{matyka2008tortuosity},
\begin{align}\label{ch4:tortuosity}
\tau & = 1 + a_{\tau} (1 - \phi(p)),
\end{align}
where $a_{\tau}$ is a fitting constant. Tortuosity $\tau$
is a decreasing function of $\phi$ because $d\tau / d\phi = -a_{\tau} < 0$. 
Note that as $p \rightarrow 0$ then $\phi \rightarrow a_{\phi}$,
and, therefore $\tau \rightarrow 1 + a_{\tau}(1 - a_{\phi})$.
The compressibility coefficient of tortuosity $\zeta_{\tau}$ is given by the expression,
\begin{align}\label{ch4:tortuosityComp}
\zeta_{\tau}(p) & = \frac{d}{dp} \ln\tau = -a_{\tau} \frac{\phi}{\tau} \zeta_{\phi}(p).
\end{align}


The turbulence correction factor, $B$ (which is a scalar in one-dimension),  is modelled by, \citet{zhang2013numerical}, \citet{thauvin1998network} and \citet{macini2011laboratory},
\begin{align}\label{ch4:TurFactor1}
B(\phi, K_a, \tau) & = \frac{a_{B} \tau^{b_{B}}}{K_a^{c_{B}}\phi^{d_{B}}}
\end{align}
where, $a_{B}$, $b_{B}$, $c_{B}$ and $d_{B}$ are empirical constants.
The Compressibility coefficient of turbulence factor $\zeta_{b}$ is given by,
\begin{align}\label{ch4:CompCoTurFactor}
\zeta_{B}(p) & = \frac{d}{dp} \ln B = b_{B} \zeta_{\tau}(p) - c_{B} \zeta_{K_a}(p)
                                             - d_{B} \zeta_{\phi}(p).
\end{align}


The control factor, $F$, introduced in Section 3 is given by
\begin{align}\label{ch4:ControlFactor1}
F = \left[ 1 + \frac{B\rho}{\mu} K_a |u| \right] ^{-1}
\end{align}
The compressibility coefficient for the control factor $\zeta_{F}$ is given by,
\begin{align}\label{ch4:CompCoControlFactor}
 \zeta_{F}(p)    &= (F-1) \left[ \zeta_{\rho}(p) - \zeta_{\mu}(p)
                           + \zeta_{K_a}(p) + \zeta_{B}(p) + \zeta_{|u|}(p) \right],
\end{align}
where $\zeta_{K_a} (p) = \zeta_{K}(p) + \zeta_{f} (p)$, and $\zeta_{K}(p)$ is given by  equation \eqref{ch4:KCComp} and $\zeta_{f}(p)$ is defined in equation \eqref{ch3:npm22}.

\subsection{Gas adsorption isotherm}
The shale adsorbs a portion of the gas, the amount retained is a dynamic process depending upon
the pressure and the adsorbant, e.g. kerogen, in the shale. As the pressure in the shale decreases,
the gas is desorbed in to the pore network.
The relationship between  pressure and the volume  of the adsorbed gas
is described by the desorption isotherm. The most commonly used relation to estimate
the amount of gas released is  the Langmuir Isotherm Formula, \citet{foo2010insights}.
\citet{cui2009measurements}, and \citet{civan2011shale} used the following formula, which is based
upon the Langmuir adsorption isotherm,
\begin{align}\label{ch4:LangAdIsotherm}
   q_a & = \frac{q_L p}{p_L + p}, \\
   q & = \frac{\rho_s M_g}{V_{std}}q_a
\end{align}
where $\rho_s$ (kg/m$^3$) denotes the material density of the porous sample,
$q$ (kg/m$^3$) is the mass of gas adsorbed per solid volume, $q_a$ (std m$^3$/kg)
is the standard volume of gas adsorbed per solid mass, $q_L$ (std m$^3$/kg) is the
Langmuir gas volume, $V_{std}$ (std m$^3$/kmol) is the molar volume of gas at
standard temperature (273.15K) and pressure (101,325Pa), $p$ (Pa) is the gas pressure,
$p_L$ (Pa) is the Langmuir gas pressure, and $M_g$ (kg/kmol) is the molecular weight
of gas. (Half of the gas molecules occupies the empty spaces along the surface of the pores at
the Langmuir pressure $p_L$.)

The compressibility coefficient of the adsorbed gas $\zeta_{q}$ is given by,
\begin{equation}\label{ch4:CompCoAdGas}
\zeta_q(p) = \frac{p_L}{p(p_L+p)}.
\end{equation}
$\zeta_q(p)$ decreases with increasing pressure. It implies that $\zeta_q(p)$ is limited by the number of empty spaces
available at the pore surface.


\subsection{Compressibility Coefficient of Knudsen number}\label{knudsen}

The compressibility coefficient of the mean free path $\zeta_{\lambda}$ is given by
\begin{equation}\label{ch4:CompCoMFP1}
\zeta_{\lambda}(p) = \zeta_{\mu}(p) - \zeta_{\rho}(p).
\end{equation}

Substituting equations \eqref{ch4:KozenyCarmanEq1} and \eqref{ch4:tortuosity} in to
equation \eqref{ch2:rad}, we obtain
\begin{equation}\label{ch4:Radius2}
R_h = 2\Gamma_{KC} \sqrt{2 + 2a_{\tau}(1-\phi)} \left( \frac{\phi}{\alpha_{KC} - \phi} \right)^{\beta_{KC}}.
\end{equation}
Note that $R_h$ increases with increase in pressure.

The compressibility coefficient of the hydraulic radius, $\zeta_{R}$, is given by, 
\begin{equation}\label{ch4:CompCoRadius}
\zeta_{R}(p) = \frac{d(\ln R_h)}{dp}= \frac{1}{2} \left[ \zeta_{\tau}(p) + \zeta_{K} - \zeta_{\phi}(p) \right].
\end{equation}
In general, $\zeta_R$ decreases with the increase in pressure.


A formula for the Knudsen number is obtained by substituting equations  \eqref{ch4:MFP1} and
\eqref{ch4:Radius2} into equation \eqref{ch4:KnudsenNumber1},
\begin{equation}\label{ch4:KnudsenNumber2}
K_n = \frac{1}{2\Gamma_{KC}} \sqrt{\frac{\pi}{2R_gT}} \frac{\mu}{\rho \sqrt{2+2a_{\tau}(1-\phi)}}
       \left( \frac{\alpha_{KC} - \phi}{\phi} \right)^{\beta_{KC}}
\end{equation}
The compressibility coefficient of Knudsen Number $\zeta_{K_n}$ is given by
\begin{align}\label{ch4:CompCoKn}
\nonumber \zeta_{K_{n}}(p) &= \frac{d(\ln K_n)}{dp} = \zeta_{\lambda} (p) - \zeta_{R} (p) \\
                           & = \zeta_{\mu}(p) - \zeta_{\rho}(p) + \frac{1}{2} \left[ \zeta_{\phi}(p) - \zeta_{\tau}(p) - \zeta_{K}(p) \right].
\end{align}
$\zeta_{K_n}$ has negative values.


\subsection{Intrinsic permeability and apparent permeability}
Using equations \eqref{ch2:aper}, \eqref{ch2:pcf}, and  \eqref{ch2:rare}, the compressibility
coefficient for $K_a$, is given by
\begin{equation}\label{ch4:RareFac1}
 \zeta_{K_a} = \zeta_{K}+ \zeta_{f}
\end{equation}
$\zeta_{K}$ is given by equation \eqref{ch4:KCComp}. To compute $\zeta_{f}$, from equation \eqref{ch2:pcf}
we have
\begin{align}
\zeta_{f} &= \frac{d(\ln(f))}{dp} =
K_n\frac{d(\ln(f))}{dK_n} \frac{1}{K_n}\frac{dK_n}{dp} = \zeta_{K_n} K_n\frac{d(\ln(f))}{dK_n}
\end{align}
where,
\begin{align}
\nonumber K_n\frac{d(\ln(f))}{dK_n} & = \left( \frac{ \sigma K_n}{ 1 + \sigma  K_n}\right)
             \left(1 + \frac{K_n}{\sigma}\frac{d\sigma}{dK_n}\right)  \\
             & \ \  + \frac{(4-b_{SF})K_n}{(1 + (4-b_{SF})K_n)}
               + \frac{b_{SF}K_n}{ 1 - b_{SF}K_n}
\end{align}
In the limit of continuous (Darcy) flow,  $K_n \rightarrow 0$, then
$f \rightarrow 1$ and $K_a \rightarrow K$. On the other hand,
in the limit of Knudsen diffusion,  $K_n \rightarrow \infty$, then $\sigma \rightarrow \sigma_0$,
$f \rightarrow \infty$ and hence $K_a \rightarrow K$. Thus in the two extreme
limits we have $K_a \rightarrow K$.

\section{Numerical Procedure}\label{numproc}

The new transport model contains 13 parameters excluding the
Forchheimer correction term, and 17 parameters if the Forchheimer correction term is included.
All of these parameters must be specified as input to the simulations. With these specified, the
correlations for the model parameters and their associated compressibility parameters are completely
determined as a function of the local pressure. The system can thus be solved numerically,
given appropriate boundary and initial conditions.

We have developed an implicit finite volume solver for the general one-dimensional system.
The balance equation \eqref{ch3:npm33} is discretised and integrated over a typical control
volume, to yield a system of nonlinear algebraic equations for the pressure field vector
$P_n=\{p_1,p_2, ..., p_N\}$, where $p_i$ is the pressure at the centre of control volume labeled
$i$, at some given time step $n$. Assuming that the system has been solved at time $t=ndt$,
then we solve the following system at $t=(n+1)dt$,
\begin{align}
     A(P)P &= S(P)
\end{align}
This is nonlinear algebraic system, so it is solved iteratively by first linearising the matrix of 
coefficients $A$ and the source term $S$ at the current iteration step, i.e. $A^\nu=A(P^\nu)$, 
and $S^\nu=S(P^\nu)$. Initially, for $\nu=0$, we have $P^0=P_n$, so that $A^0=A(P^0)$, 
and $S^0=S(P^0)$. Then we solve for $P^{\nu+1}$, until convergence,
\begin{align}
     A^\nu P^{\nu+1} &= S^\nu, \\
                P^{\nu+1} &= (A^\nu)^{-1}S^\nu \\
                      \hbox{define}\ \epsilon^{\nu+1} &=  \frac{1}{N}\sum_{i=1}^N
                                               \left|{\frac{p_i^{\nu+1}-p_i^\nu}{p_i^{\nu+1}}}\right| \\
\nonumber     \hbox{if}\  \epsilon^{\nu+1}&< \epsilon_{Tol} \implies  \hbox{converged}\\
         \implies      &P_{n+1} =P^{\nu+1}; \hbox{ go to next time step $n \to n+1$} \\
\nonumber    \hbox{else   if}\  \epsilon^{\nu+1} &> \epsilon_{Tol} \implies  \hbox{not  converged}\\
         \implies      &\hbox{go to next iteration $\nu\to \nu+1$}
\end{align}
The matrix $A^\nu$ is tri-diagonal and can therefore be inverted easily. All the model parameters and compressibility coefficients are fully pressure dependent at all times,  and these must be updated at every iteration step $\nu$. $\epsilon_{Tol}$  is a small tolerance level for the relative error $\epsilon_{\nu}$, typically less than $10^{-4}$.

A flux limiter for large gradients in the pressure field is included. The implicit nature of the solver gives the method good stability.

The solver for a steady-state systems is the same, except that we do not advance the time step.

\section{Determining rock properties using the transport model}
\label{resim1}

In this work, the main application of the new transport model is to the problem of determining rock
properties. This is carried out through solving an inverse problem whereby model parameters
are adjusted to fit a given set of experimental data.

Many studies have been conducted to determine the permeability of shale rocks,
but in most of them the reservoir parameters were assumed to be pressure
independent or constant in application, see for example \citet{brace1968permeability},
\citet{hsieh1981transient},
\citet{neuzil1981transient}, \citet{liang2001nonlinear}, \citet{malkovsky2009new},
\citet{cui2009measurements}, and \citet{civan2011shale}.
However, it is more realistic to consider some, if not all, of the model parameters as being pressure dependent, see \citet{pong1994non}, \citet{beskok1999report},
\citet{roy2003modeling}, \citet{javadpour2009nanopores},
\citet{civan2005improved}, \citet{ali2015compressibility}.

We will use the new transport model equation \eqref{ch3:npm33} with equations \eqref{ch3:npm34}
and \eqref{ch3:npm34a}  in order to estimate rock core properties of rock samples
for which we have experimental data from the pressure-pulse decay tests of \citet{pong1994non},
who used Nitrogen as the working gas. See also, \citet{civan2011shale},
\citet{lorinczi2014direct}, \citet{wang2015variations},
\citet{huang2015applications}, and \citet{ali2015compressibility}.
Pong's data-sets consist of measurements of pressure, $p$, at a number of stations, $x$,
along the rock length; this is repeated for several different inlet pressures, $P_{in}$.

\subsection{Pressure pulse-decay tests}
\begin{figure}[t]
	\centering
		\includegraphics[width=10cm]{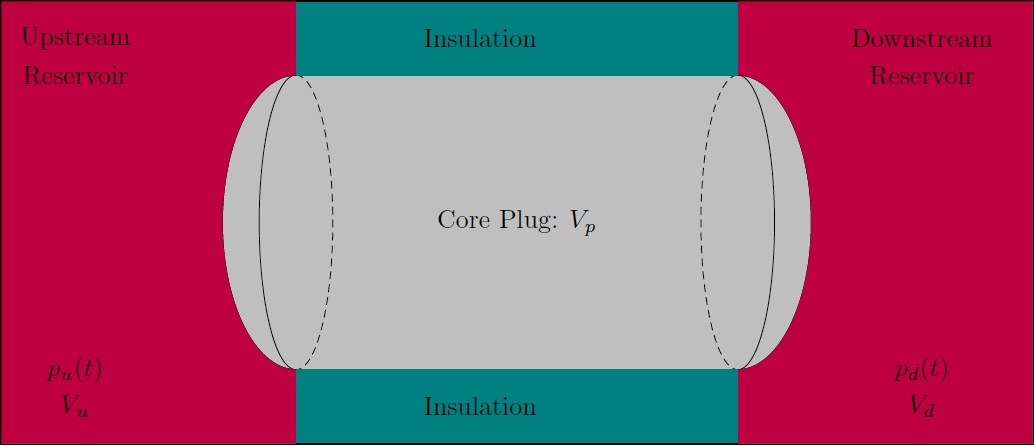}
    \caption{Pressure-pulse decay test setup}
	\label{fig:PPDT2}
\end{figure}

In a pressure-pulse decay test, Fig. 5, a short rock sample  of length $L$ is initially at a constant pressure
inside the core sample itself. A pulse of pressure is then sent through the sample from the upstream
boundary and the pressure field quickly reaches a steady state distribution across the core length. The
pressure is recorded at different stations along the core length.  Mathematically, we have the following
initial condition,
  \begin{equation} \label{ch3:npm48}  
p(x,0) = p_o, \qquad 0\leq x \leq L, \quad t=0,
\end{equation}
and Dirichlet Boundary Conditions,
\begin{align} \label{ch3:npm49} 
p(0,t)=p_u(t), \qquad  x=0, \quad t>0.\\
p(L,t)=p_d(t), \qquad  x=L, \quad t>0.
\end{align}
where $p_u$ is the measured values of the pressure in the upstream reservoir, 
and $p_d$ is the measured values of the pressure in the downstream reservoir.

The flux  conditions at the inlet (upstream) boundary is
$\frac{d(\rho V_u)}{dt}=-\rho \textbf{u}\cdot\textbf{n}A$, which for one
dimensional domain becomes,
\begin{equation}\label{ch3:npm52}  
\frac{\partial p}{\partial x} =  \left[ \frac{ V_u\mu\phi L}{V_p} \frac{\zeta_{\rho}(p)}{F K_a}\right]
\frac{\partial p}{\partial t},  \qquad p=p_u;\quad x=0, \quad t>0.
\end{equation}
where $V_u$ is the volume of upstream reservoir, and $A$ is the cross-sectional
area.

The flux  conditions at the outlet (downstream) boundary is
$\frac{d(\rho V_d)}{dt}=\rho \textbf{u}\cdot\textbf{n}A$, which for one
dimensional domain becomes,
\begin{equation}\label{ch3:npm54}  
\frac{\partial p}{\partial x} = - \left[ \frac{ V_d\mu\phi L}{ V_p} \frac{\zeta_{\rho}(p)}{F K_a} \right]
\frac{\partial p}{\partial t},  \qquad p = p_d; \quad x = L, \quad t>0.
\end{equation}
where $V_d$ is the volume of downstream reservoir.

In the pressure-pulse decay tests of \citet{pong1994non}, several different experiments with different
inlet pressures $P_{in}$ were carried out -- the values of $P_{in}$ chosen are listed in Table \ref{Table01}.

\subsection{Sixteen transport  models}\label{methodology}

How important is it to keep all parameters to be pressure dependent at all times? To address this
question, we consider sixteen different transport models, labeled $k = 1, 2,
\cdots, 16$. These models are produced by taking the four basic compressibility coefficients that
appear in the model, ($\zeta_\rho, \zeta_K, \zeta_f, \zeta_\mu$), to be pressure-dependent or
pressure-independent -- this gives 16 combinations resulting in the different transport models,
listed in Table \ref{Table02}.

Model 1, is when all the parameter are pressure independent, and all compressibility coefficients
are zero, $\zeta_\rho=\zeta_K=\zeta_f=\zeta_\mu=0$. This collapses to the Darcy law.

Model 16 is the fully pressure dependent case,  $\zeta_\rho\not=0, \zeta_K\not=0, \zeta_f\not=0, \zeta_\mu\not=0$.

An initial choice of model parameter values must be made, and these are then adjusted to yield the best-fit
choice of model parameters values. The rock properties $K$, $\phi$, and $\tau$ are determined as the best fit among all the models considered.


\begin{table}[t]
\centering
\addtolength{\tabcolsep}{-2pt} 
		\begin{tabular}{p{3cm} p{3cm}}
        \multicolumn{2}{c}{Civan's Inflow Conditions} \\
\# &  $p_{\text{inlet}}$ \\
 $1$  & {135}  kPa  \\
 $2$  & {170}  kPa  \\
 $3$  & {205} kPa  \\
 $4$  & {240} kPa  \\
 $5$  & {275}  kPa  \\
\end{tabular}	
\vskip 0.3cm
\caption[Civan's inflow condition]{ Inflow conditions for Civan's steady state case.}
\label{Table01}
\end{table}


\begin{table}[t]
\centering
	\begin{tabular}{p{1.5cm}p{1cm}p{1cm}p{1cm}p{1cm} p{2cm}}
	\hline\noalign{\smallskip}
                   &Compressibility & &coefficient& & \\
    Model \#  & $\zeta_{\rho}$ & $\zeta_{K}$ & $\zeta_{f}$ & $\zeta_{\mu}$ & Error \\
	\hline\noalign{\smallskip}
    1 & 0 & 0 & 0 & 0 &  2.69e-02\\
	2 & $p$ & 0 & 0 & 0 & 2.68e-02 \\
      3 & 0 & $p$ & 0 & 0 &  1.64e-01 \\
	4 & 0 & 0 & $p$ & 0 & 1.00e+05 \\
      5 & 0 & 0 & 0 & $p$ &  2.69e-02 \\
	6 & $p$ & $p$  & 0 &  0 & 2.23e+00 \\
      7 & $p$ & 0 & $p$ & 0 &  8.52e-01 \\
	8 & $p$  & 0 & 0 & $p$  &  2.69e-02 \\
      9 & 0 & $p$ & $p$ & 0 &  1.19e+00 \\
    10 & 0 &  $p$  & 0 &  $p$  &  1.64e-01 \\
    11 & 0 & 0 & $p$  & $p$ &  1.00e+05 \\
    12 & $p$  &  $p$  & $p$  & 0 & 1.059e-04 \\
    13 & $p$  &  $p$  & 0 & $p$ & 2.23e+00 \\
    14 & $p$ & 0 & $p$  & $p$ &  8.52e-01 \\
	15 & 0 & $p$  & $p$  & $p$  & 1.19e+00 \\
	16 & $p$ & $p$ &  $p$  & $p$  & 1.055e-04 \\
	\noalign{\smallskip}\hline\noalign{\smallskip}
	\end{tabular}
\vskip 0.1cm
\caption[Sixteen Cases]{List of simulations carried out. The four basic compressility coefficients
    are made pressure-dependent or pressure-independent: in columns 2-5, an entry of '0' means
    that the compressibility factor is zero, and an entry of '$p$' means that it is nonzero and
    the associated physical parameter is function of pressure $p$. Column 6 shows the error from the
    simulations (with $B=0$), using equation \eqref{Error}.}
\label{Table02}
\end{table}

\begin{table}[t]
\centering
\addtolength{\tabcolsep}{-2pt} 
		\begin{tabular}{p{5cm} p{3cm}}
       \multicolumn{2}{c}{Civan (2011)}  \\
Reservoir  Parameters & Values      \\
$L$ (m)        & 0.003  \\
$N_x$          & 100    \\
$R_g$ ($ J/kMol/K$)     & 8314.4 \\
$M_g$ ($ Kg/KMol/K$)  & 28.013 \\
$T$ ($K$)         & 314          \\
$p_c$ ($KPa$)   & 3396          \\
$t_c$ ($K$)       & 126.19         \\
$b_{SF}$          & {-1}            \\
$\sigma_0$                  & {1.3580}       \\
$A_{\sigma}$               & {0.1780}       \\
$B_{\sigma}$                & {0.4348}      \\
$\tau $                    & {1}            \\
$\zeta_{\tau}$ $(Pa^{-1})$  & {0}   \\
$\phi$                    & {0.2}           \\
$\zeta_{\phi}$  $(Pa^{-1})$  & {5e-6} \\
$\mu$ ($Pa-s$)  & {1.85e-5}   \\
$\zeta_{\mu}$ $(Pa^{-1})$ & {3e-11} \\
$K$                     & {1e-15}           \\
$\zeta_{K}$  $(Pa^{-1})$ & {1e-6}    \\
\end{tabular}	
\vskip 0.3cm
\caption[Civan's Data Set]
{Reservoir parameters used in Civan's Model (2011). }
\label{Table03}
\end{table}


\begin{table}[t]
\centering
\addtolength{\tabcolsep}{-2pt} 
		\begin{tabular}{p{5cm} p{3cm}}
        \multicolumn{2}{c}{New Model} \\
Reservoir  Parameters & Values \\
 $L$ (m)  & 0.003 \\
 $N_x$                    & 100 \\
 $R_g$ (J/KMol/K)    & {8314.4}  \\
$M_g$ (Kg/KMol/K)  & {28.013} \\
  $T$ (K)   & {314}  \\
 $p_c$ (KPa)            & {3396} \\
 $t_c$ (K)                & {126.19} \\
 $b_{SF}$                  & {-1}   \\
$\sigma_0$                & {1.3580} \\
 $A_{\sigma}$           & {0.1780} \\
 $B_{\sigma}$            & {0.4348}\\
 $a_{\tau}$                & {1e-6} \\
 $a_{\phi}$                & {0.15} \\
 $b_{\phi}$                & {-0.939e-06} \\
$c_{\phi}$                 & {2.2} \\
$\alpha_{KC}$            & {1} \\
 $\beta_{KC}$            & {0.9}  \\
 $\Gamma_{KC}$       & {1.9e-9}\\
 \text{tol}                   & {1e-6} \\
\end{tabular}	
\vskip 0.3cm
\caption[Test Data Set]{Reservoir parameters used in new transport
model \eqref{SSM1}. A data set of base values is obtained which will
be used for  further data analysis and parameter estimation.}
\label{Table04}
\end{table}



\subsection{The general form of $\zeta_3(p)$}\label{SSMvalid}


From equation \eqref{ch3:npm19}, the most general form of $\zeta_3$ in a one-dimensional domain is,
\begin{equation}\label{SSM2a}
 \zeta_3(p) = F \left[ \zeta_{\rho}(p) +  \zeta_{K}(p) + \zeta_{f}(p) - \zeta_{\mu}(p)\right] +
               (F-1) \left[ \zeta_{b}(p) +  \zeta_{|u|}(p) \right]
\end{equation}
Without the turbulence correction
factor, i.e. $B=0$ (or $F = 1$), we obtain,
\begin{equation}\label{SSM2b}
\zeta_3(p) = \zeta_{\rho}(p) +  \zeta_{K}(p) + \zeta_{f}(p) - \zeta_{\mu}(p).
\end{equation}

In the ensuing analysis, we will first consider the simulations from models without  Forchheimer's
correction term, $B=0$ ($F=1$), and then the  simulations from models with  Forchheimer's
correction term  $B\not=0$ ($F\not=1$).


\subsection{Simulation results} \label{inparams}

Many of the model parameters are known either on physical grounds, such as the molecular weight of Nitrogen,
or by experimental set-up, such as the size of the domain, $L=3mm$. However, other model
parameters and their compressibility coefficients are modeled as described in Sections 3 and 4.

The simulation conditions matched the conditions of Pong's pressure-pulse decay tests with Nitrogen as
the working fluid, for five different inlet pressures, $P_{in}$, shown in Table \ref{Table01}.

The sixteen transport models were run with the model parameters, listed in Table \ref{Table04},
which were chosen to match Civan's parameter values, Table \ref{Table03}, where possible.
The additional parameters in the new model were initially guesstimated, and then adjusted for
best fit. Note that most of Civan's model parameters are made constant.



The relative error between the simulated results, $P^{sim}$, and the experimental data,
$P^{data}$, is determined from,
\begin{equation}\label{Error}
\epsilon^2 = \sum_{i = 1}^{N} \left| \frac{ p_{i}^{sim} - p_{i}^{data}  }{ p_{i}^{data} } \right|^2,
\end{equation}
where the summation is taken over the $N$ stations along the $x$-direction where the measurements are
recorded; here $N=6$.

\subsection{Simulation results without Forchheimer's correction ($B = 0$)}
\begin{figure}[ht]   
\begin{minipage}[b]{0.6\linewidth}
      \includegraphics[width=0.8\linewidth]{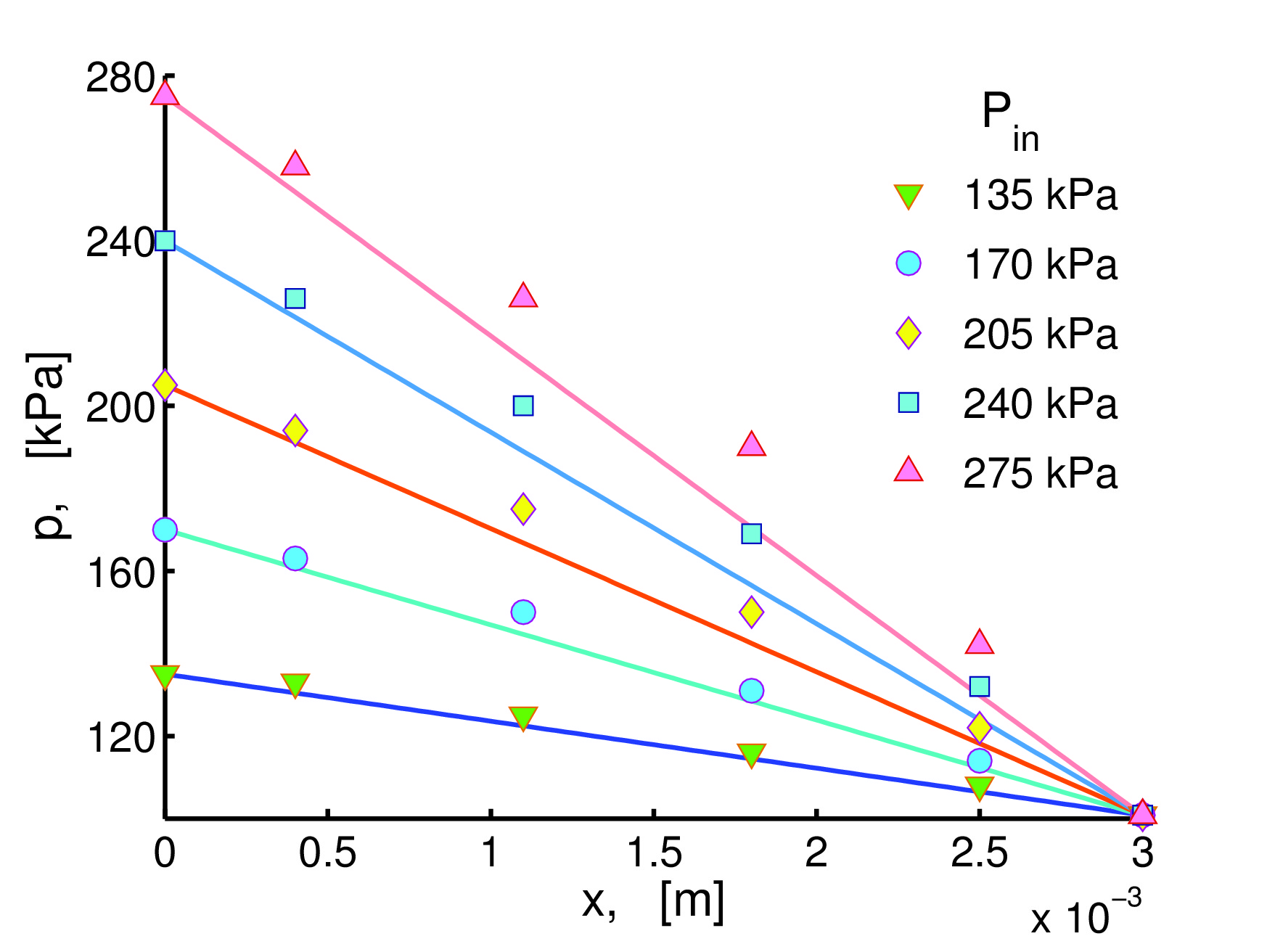}
      \captionsetup{labelformat=empty}
	\caption {Model 1}	\label{Case1PlotB}
\end{minipage}
\hspace{-1.5cm}
\begin{minipage}[b]{0.6\linewidth}
      \includegraphics[width=0.8\linewidth]{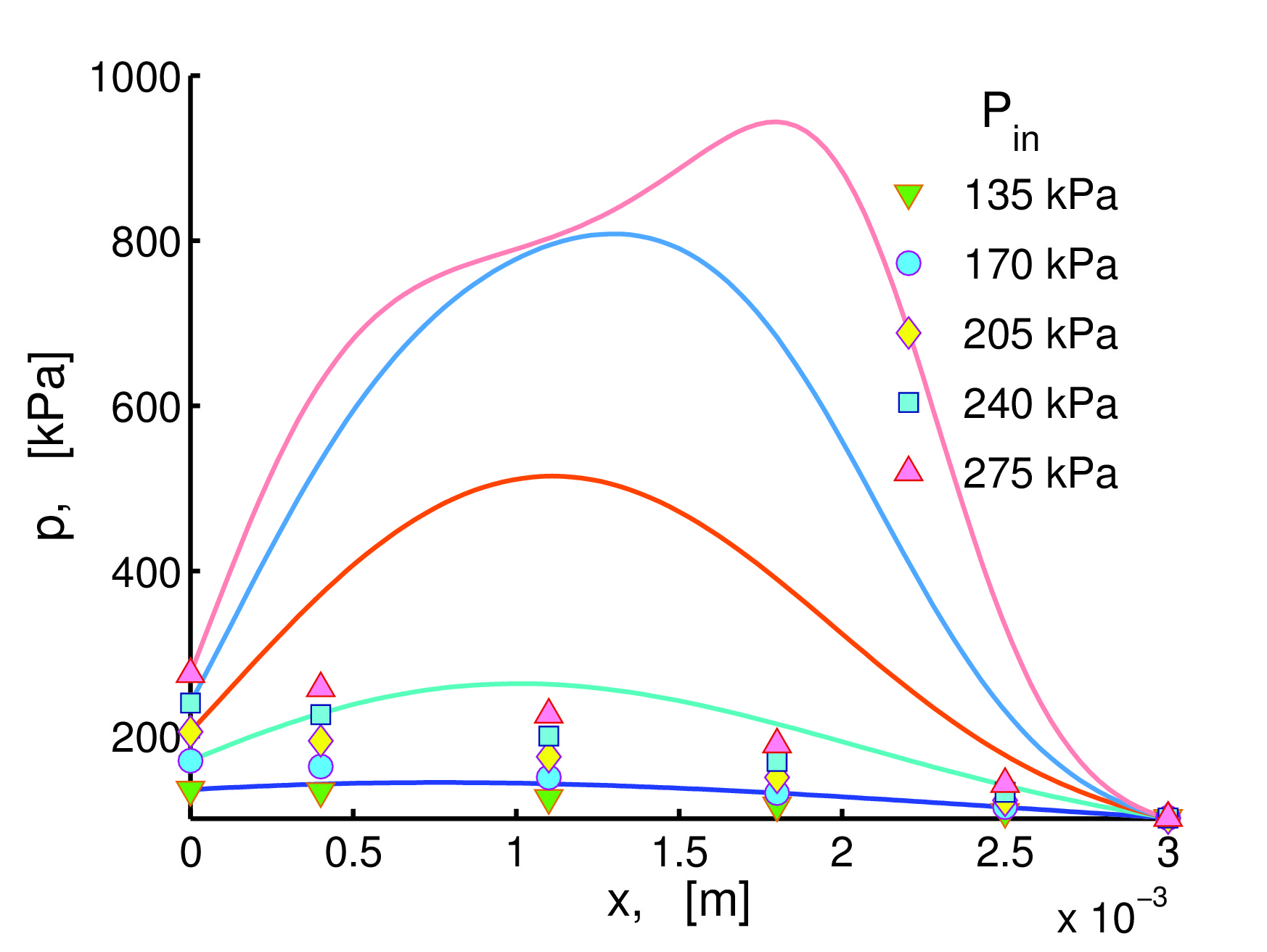}
      \captionsetup{labelformat=empty}
	\caption{Model 2} 	\label{Case2PlotA}
\end{minipage}
\begin{minipage}[b]{0.6\linewidth}
      \includegraphics[width=0.8\linewidth]{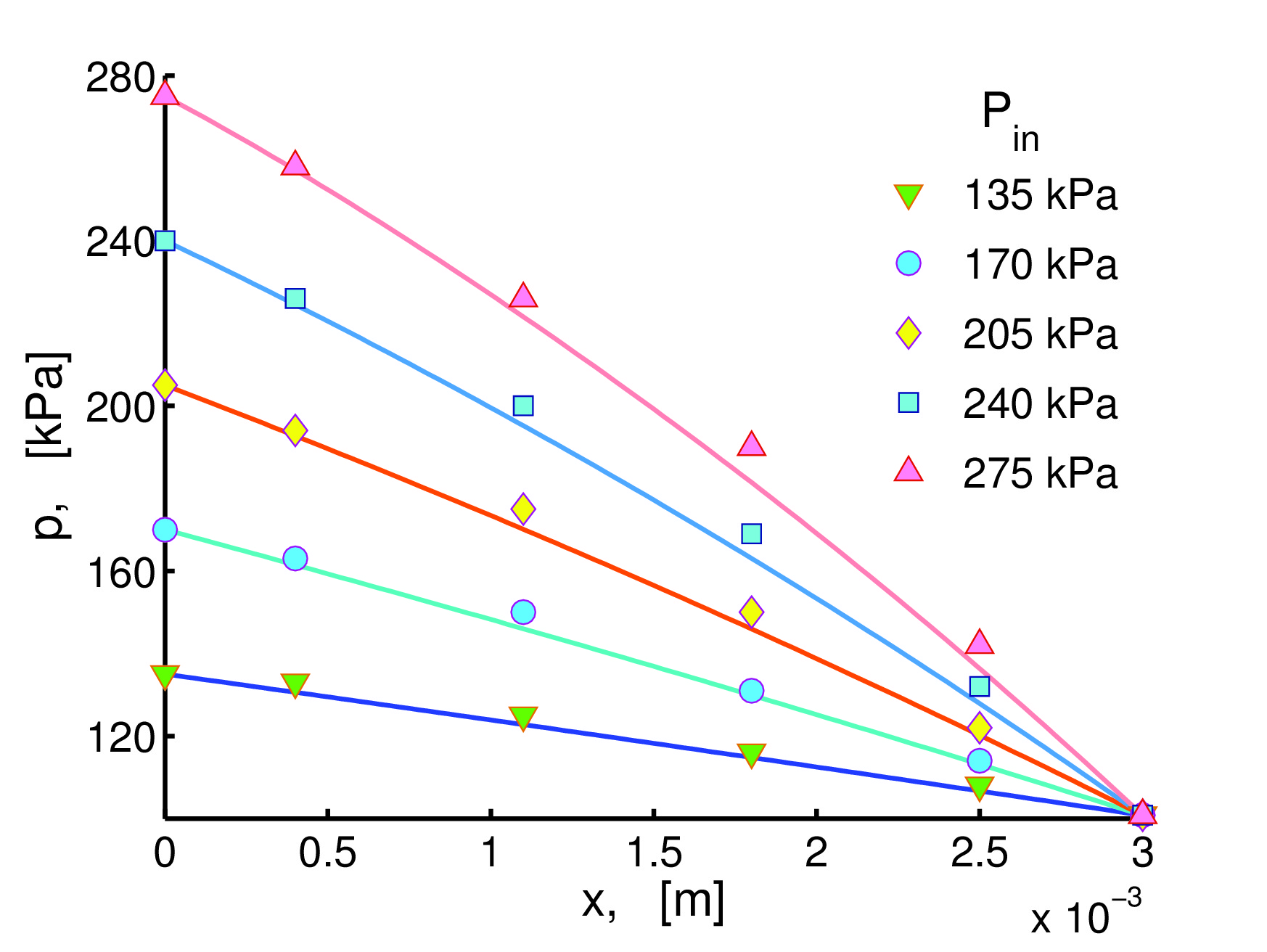}
      \captionsetup{labelformat=empty}
	\caption[]{Model 3}	\label{Case1Plot3}
\end{minipage}
\hspace{-1.5cm}
\begin{minipage}[b]{0.6\linewidth}
      \includegraphics[width=0.8\linewidth]{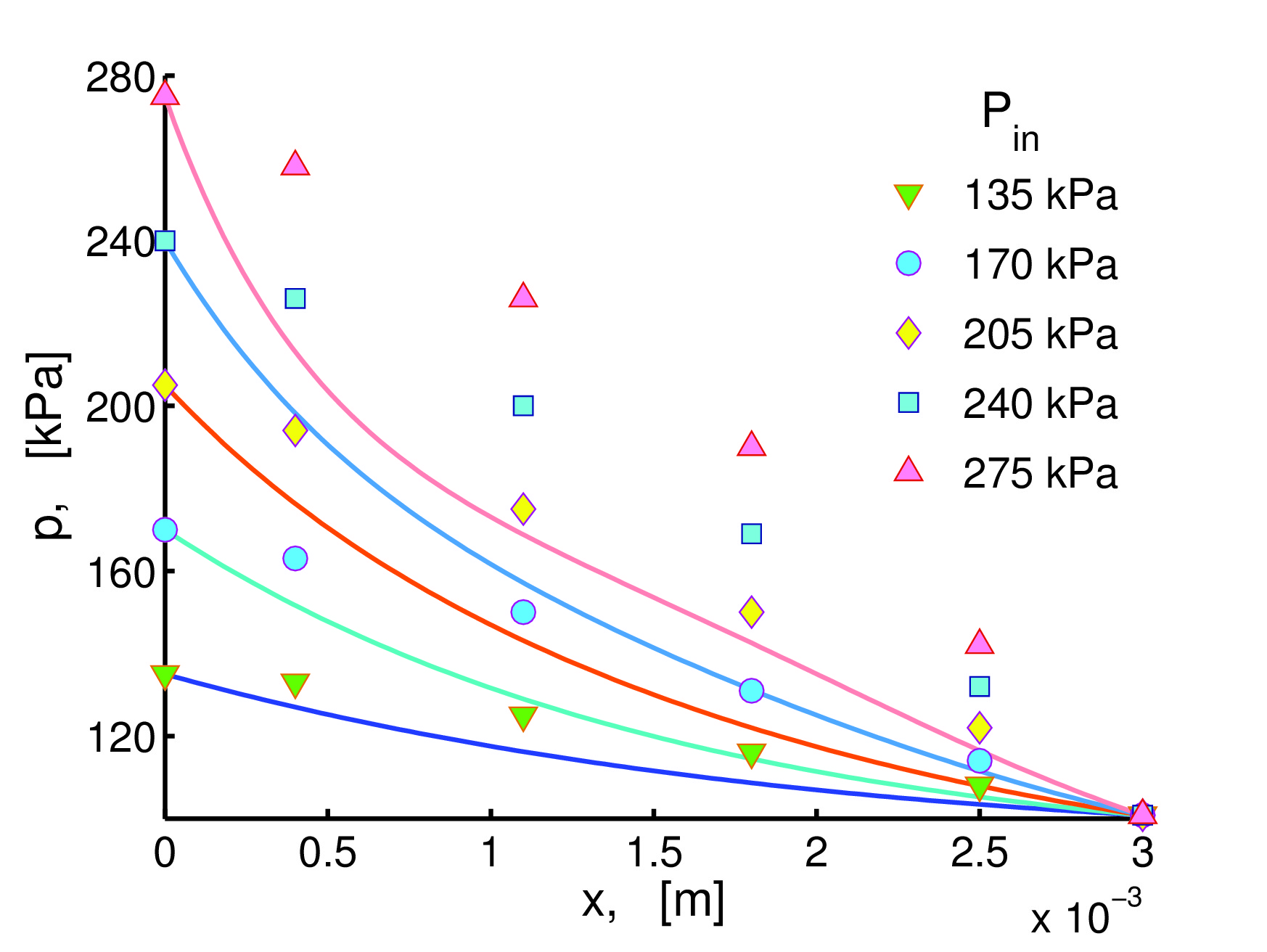}
      \captionsetup{labelformat=empty}
	\caption{Model 4} 	\label{Case2Plot4}
\end{minipage}
      \captionsetup{labelformat=empty}
	\caption{Figure 6(a). Simulation results (lines), and experimental data (symbols) from \citet{pong1994non},
               for different inlet flow pressures $P_{in}$, as indicated. Model numbers and pressure-dependent
               parameters  are shown inTable \ref{Table02}. (Similarly for Figures 6(b)-6(d).)}
\end{figure}

The simulation results, without the Forchheimer's correction term, i.e. $B=0$ ($F = 1$),  from all
sixteen models considered are shown as pressure against the distance along the core sample,
in Figures \ref{Case1PlotB}(a) - \ref{Case1PlotB}(d). The simulation results (lines) are compared
with the data of \citet{pong1994non} (symbols). The relative errors obtained are shown in Table \ref{Table02} and plotted in Figure 7.


\begin{figure}[ht]   
\begin{minipage}[b]{0.6\linewidth}
\includegraphics[width=0.8\linewidth]{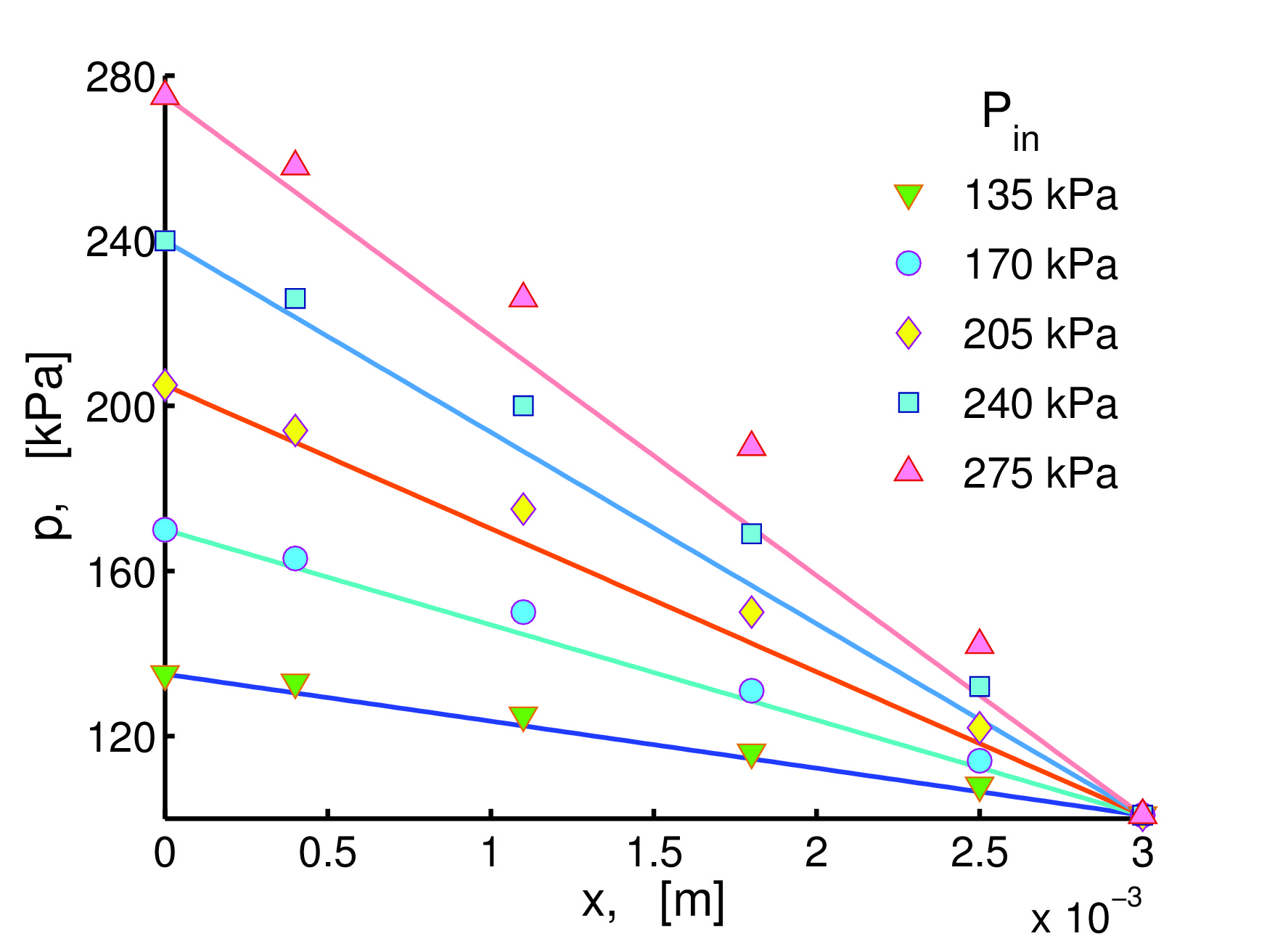}
      \captionsetup{labelformat=empty}
	\caption{Model 5} 	\label{Case1Plot5}
\end{minipage}
\hspace{-1.5cm}
\begin{minipage}[b]{0.6\linewidth}
\includegraphics[width=0.8\linewidth]{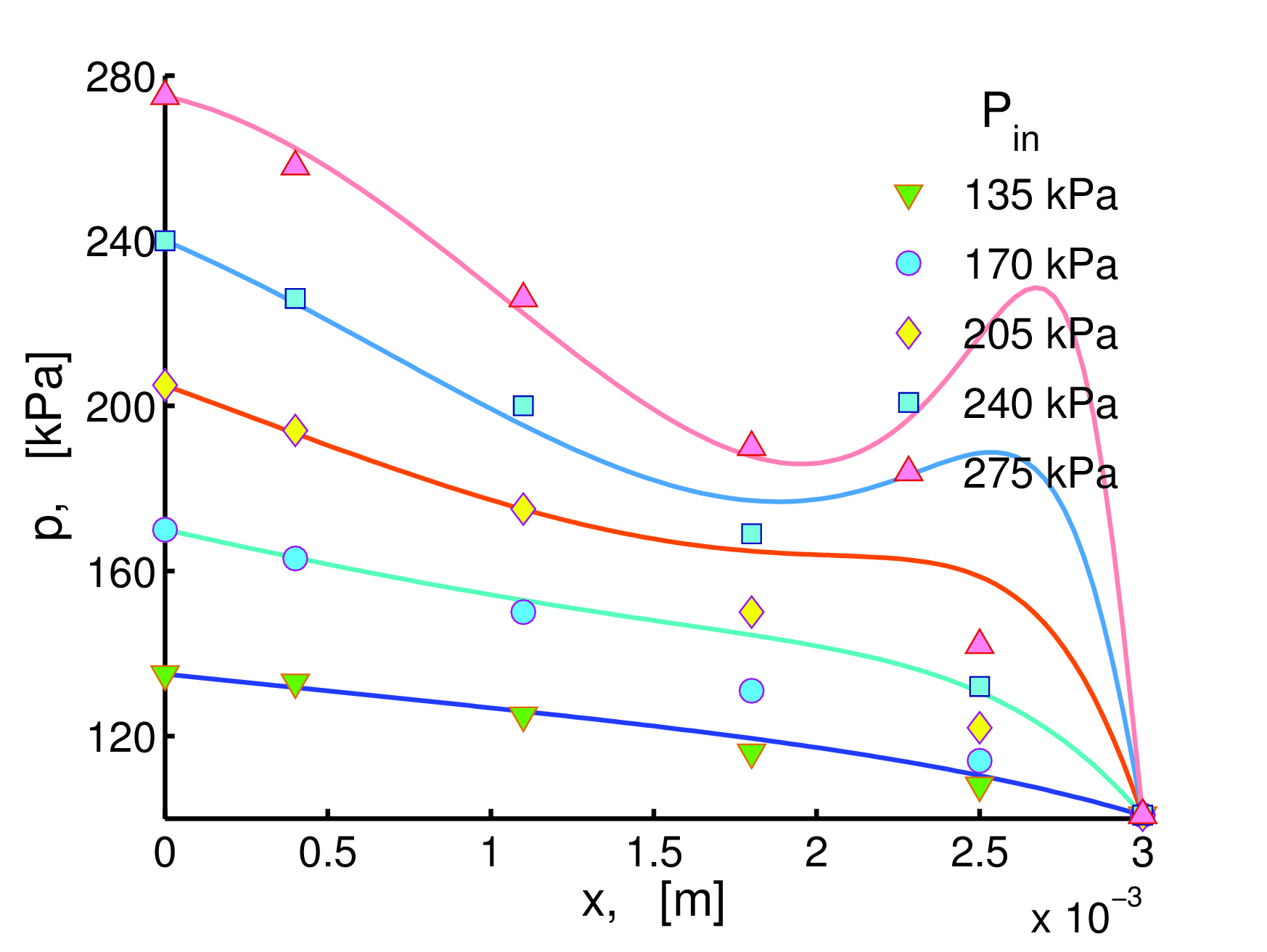}
      \captionsetup{labelformat=empty}
	\caption{Model 6}  	\label{Case2Plot6}
\end{minipage}
\begin{minipage}[b]{0.6\linewidth}
\includegraphics[width=0.8\linewidth]{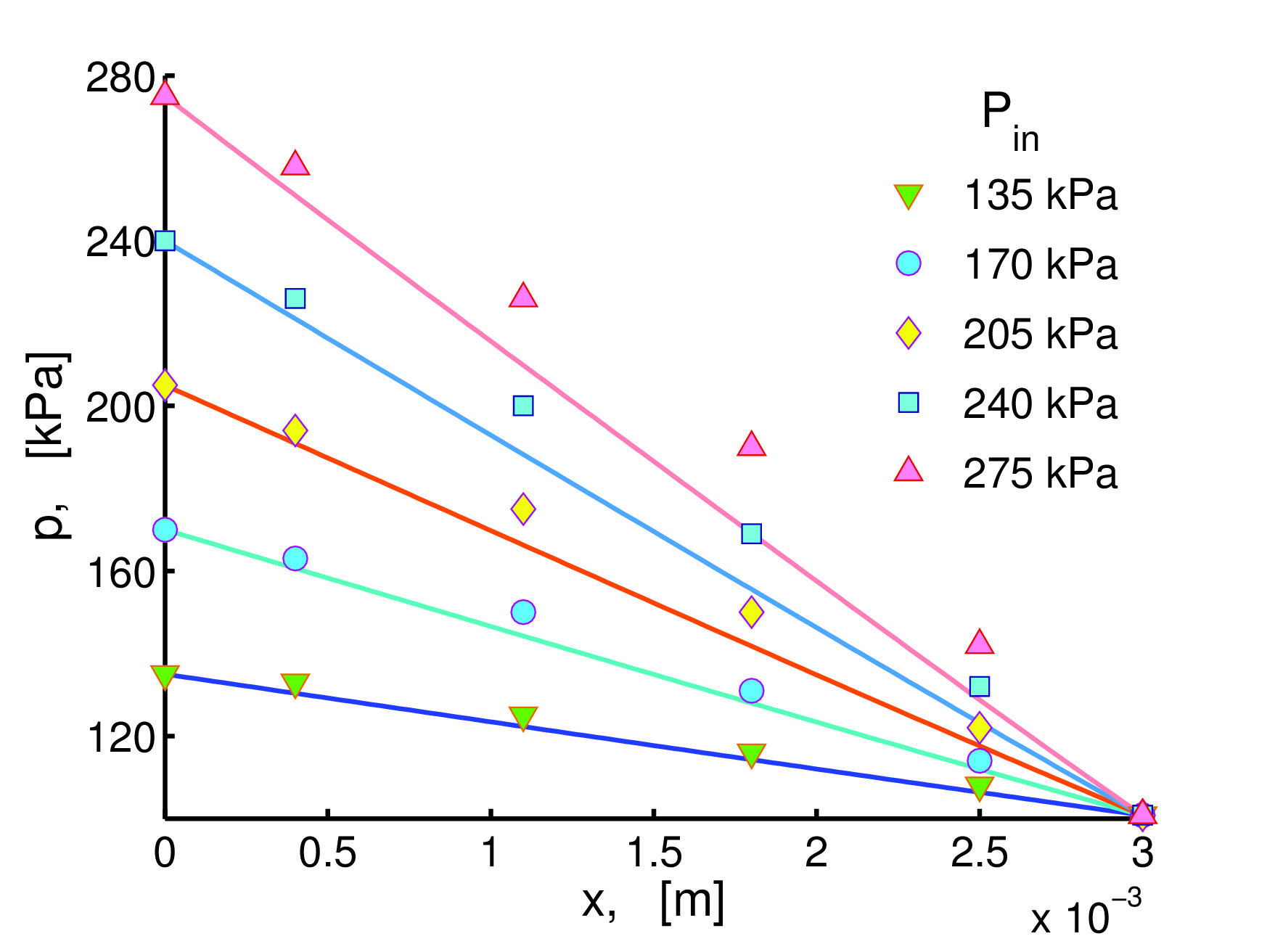}
      \captionsetup{labelformat=empty}
	\caption{Model 7} 	\label{Case1Plot7}
\end{minipage}
\hspace{-1.5cm}
\begin{minipage}[b]{0.6\linewidth}
\includegraphics[width=0.8\linewidth]{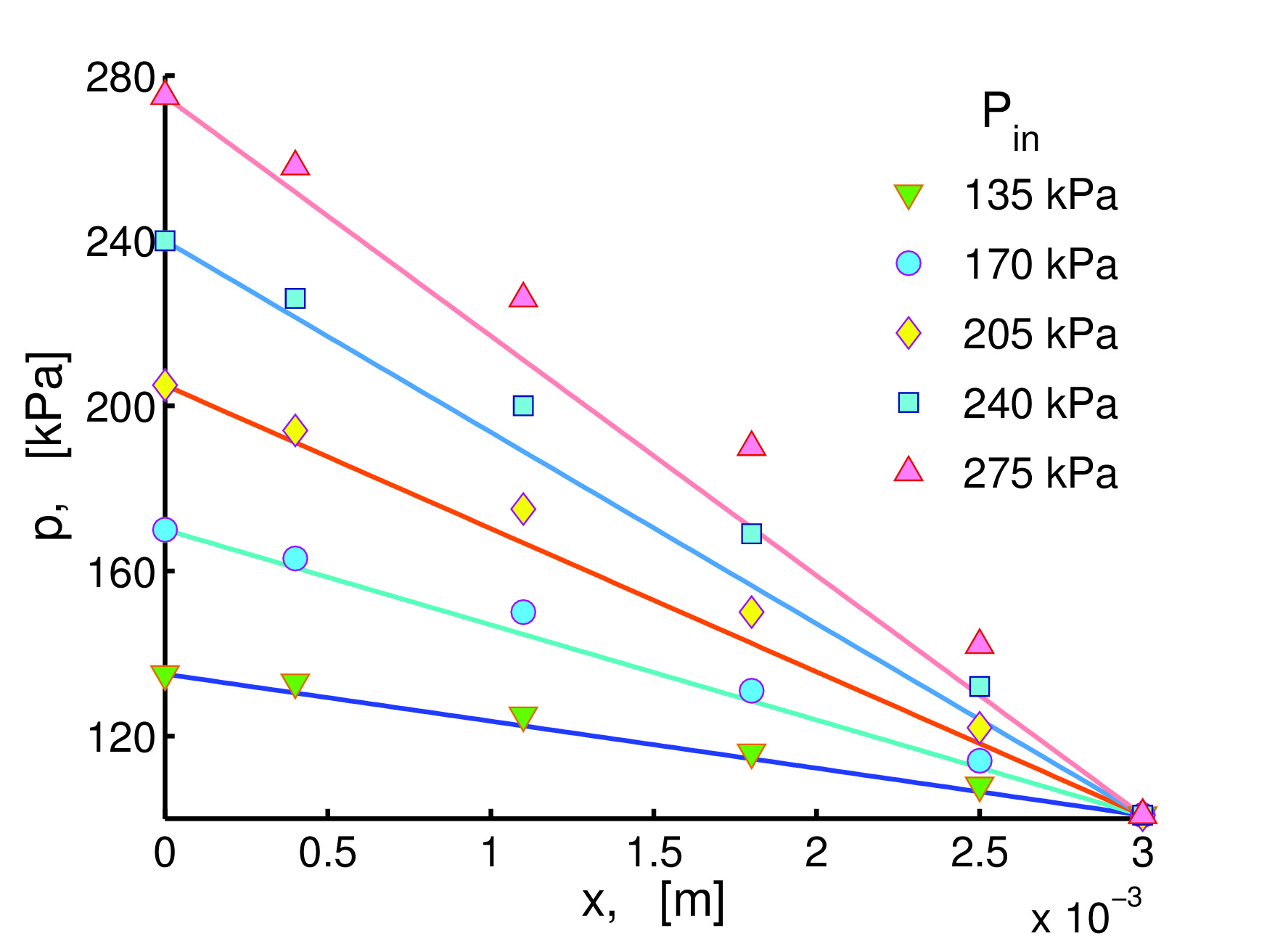}
      \captionsetup{labelformat=empty}
	\caption{Model 8}	\label{Case2Plot8}
\end{minipage}
      \captionsetup{labelformat=empty}
	\caption{Figure 6(b). (See the caption to Figure 6(a)).}
\end{figure}


\begin{figure}[ht]   
\begin{minipage}[b]{0.6\linewidth}
\includegraphics[width=0.8\linewidth]{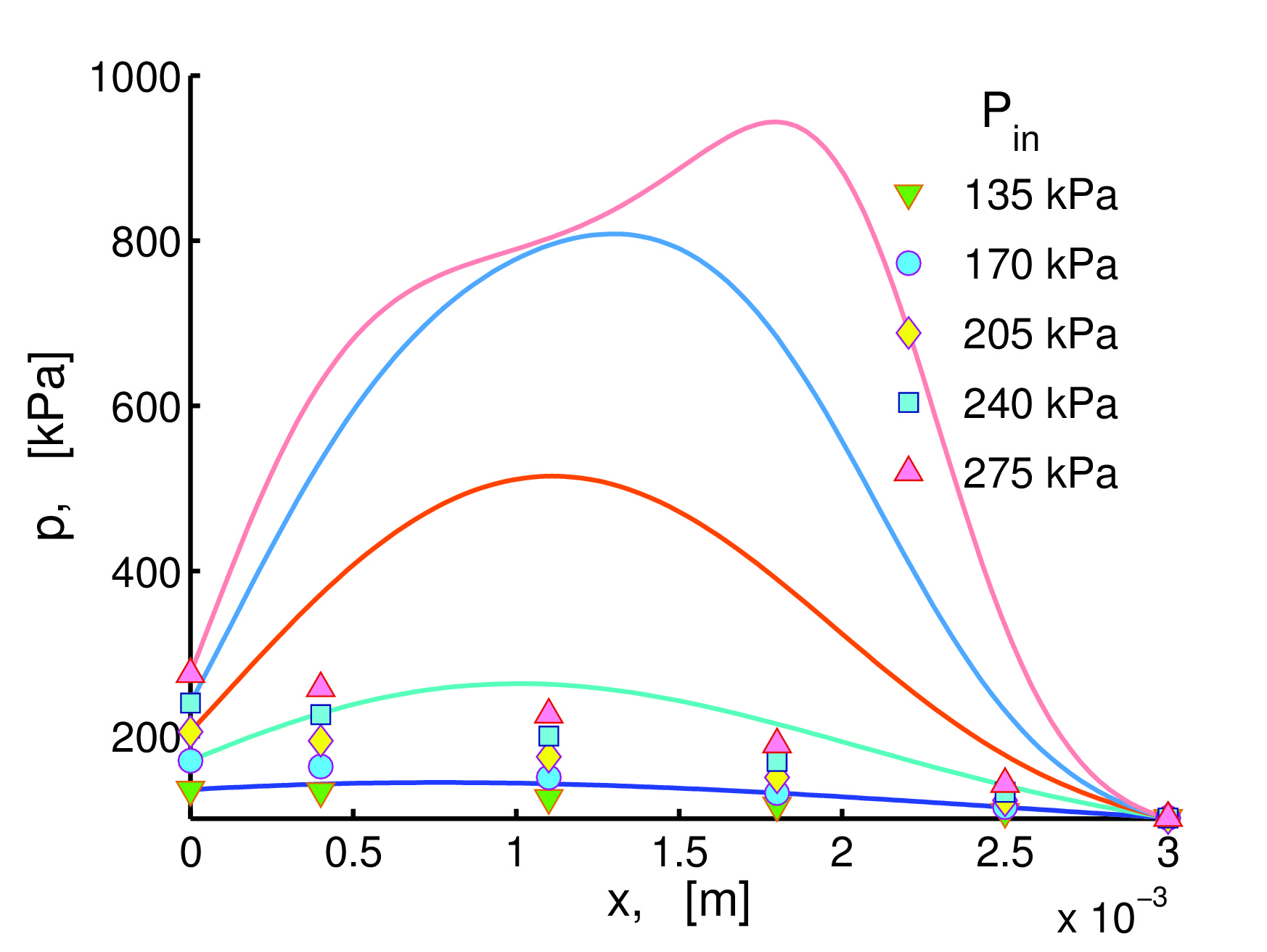}
      \captionsetup{labelformat=empty}
	\caption{Model 9} 	\label{Case1Plot9}
\end{minipage}
\hspace{-1.5cm}
\begin{minipage}[b]{0.6\linewidth}
\includegraphics[width=0.8\linewidth]{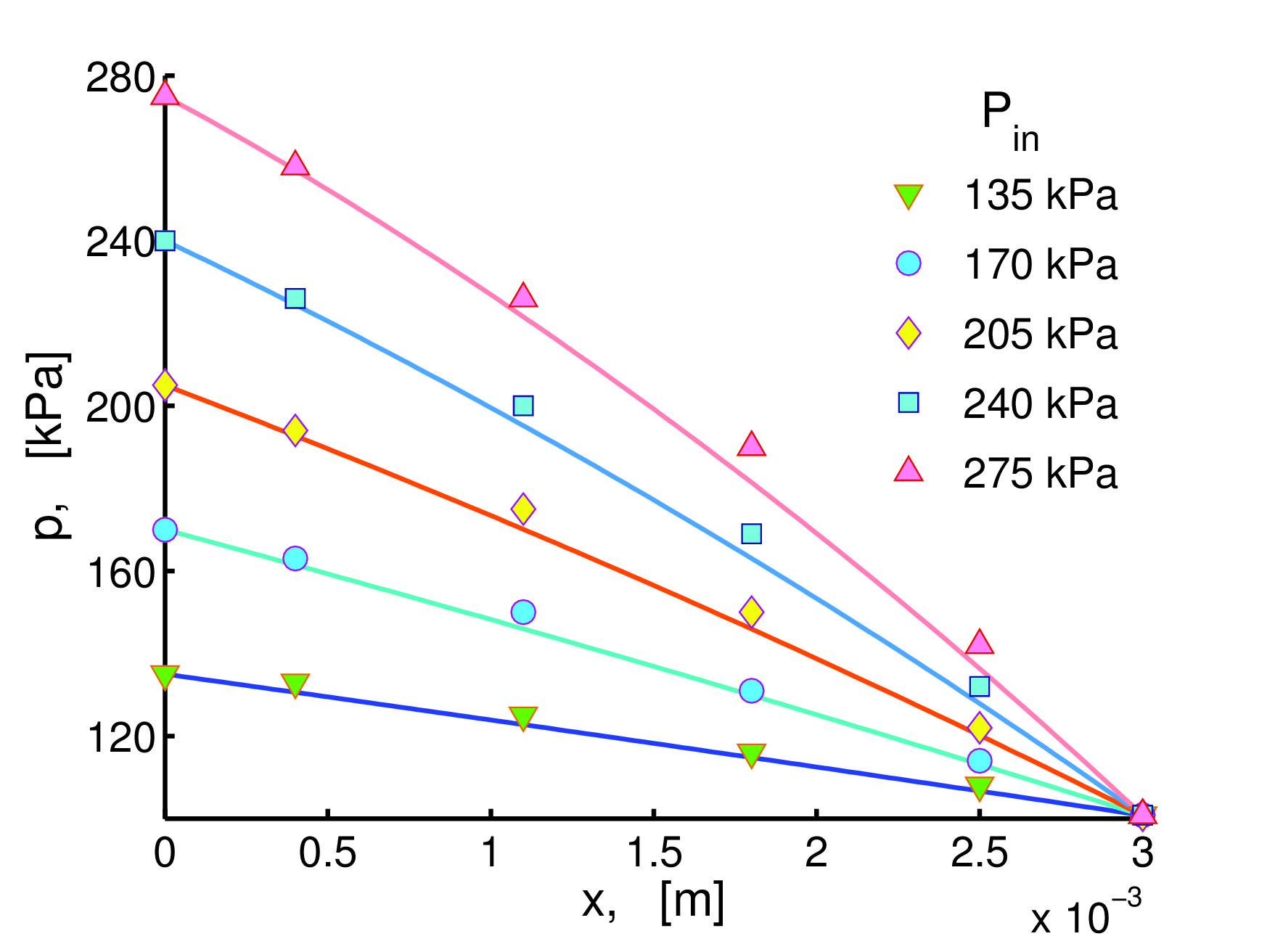}
      \captionsetup{labelformat=empty}
	\caption{Model 10}  	\label{Case2Plot10}
\end{minipage}
\begin{minipage}[b]{0.6\linewidth}
\includegraphics[width=0.8\linewidth]{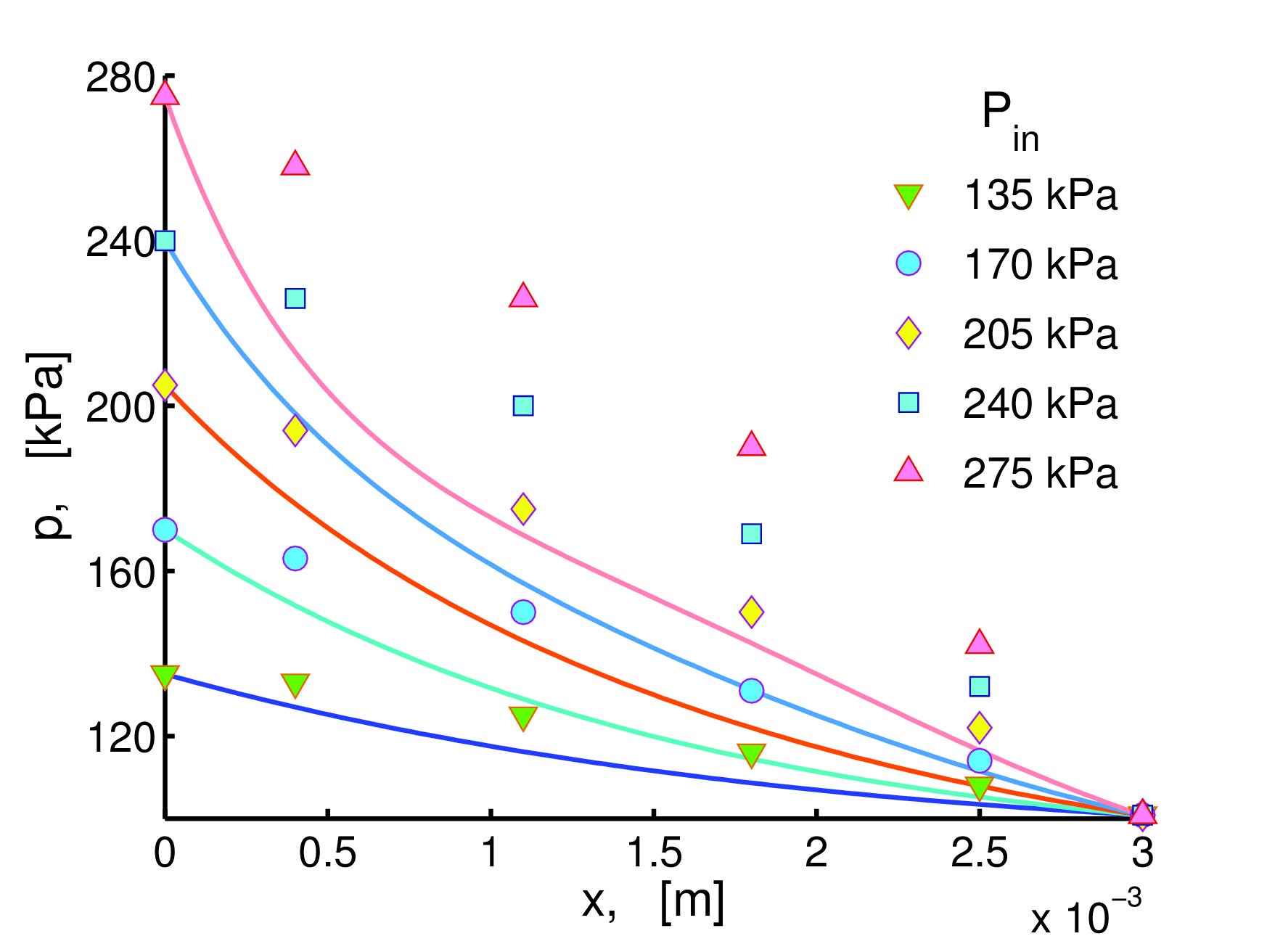}
      \captionsetup{labelformat=empty}
	\caption{Model 11} 	\label{Case1Plot11}
\end{minipage}
\hspace{-1.5cm}
\begin{minipage}[b]{0.6\linewidth}
\includegraphics[width=0.8\linewidth]{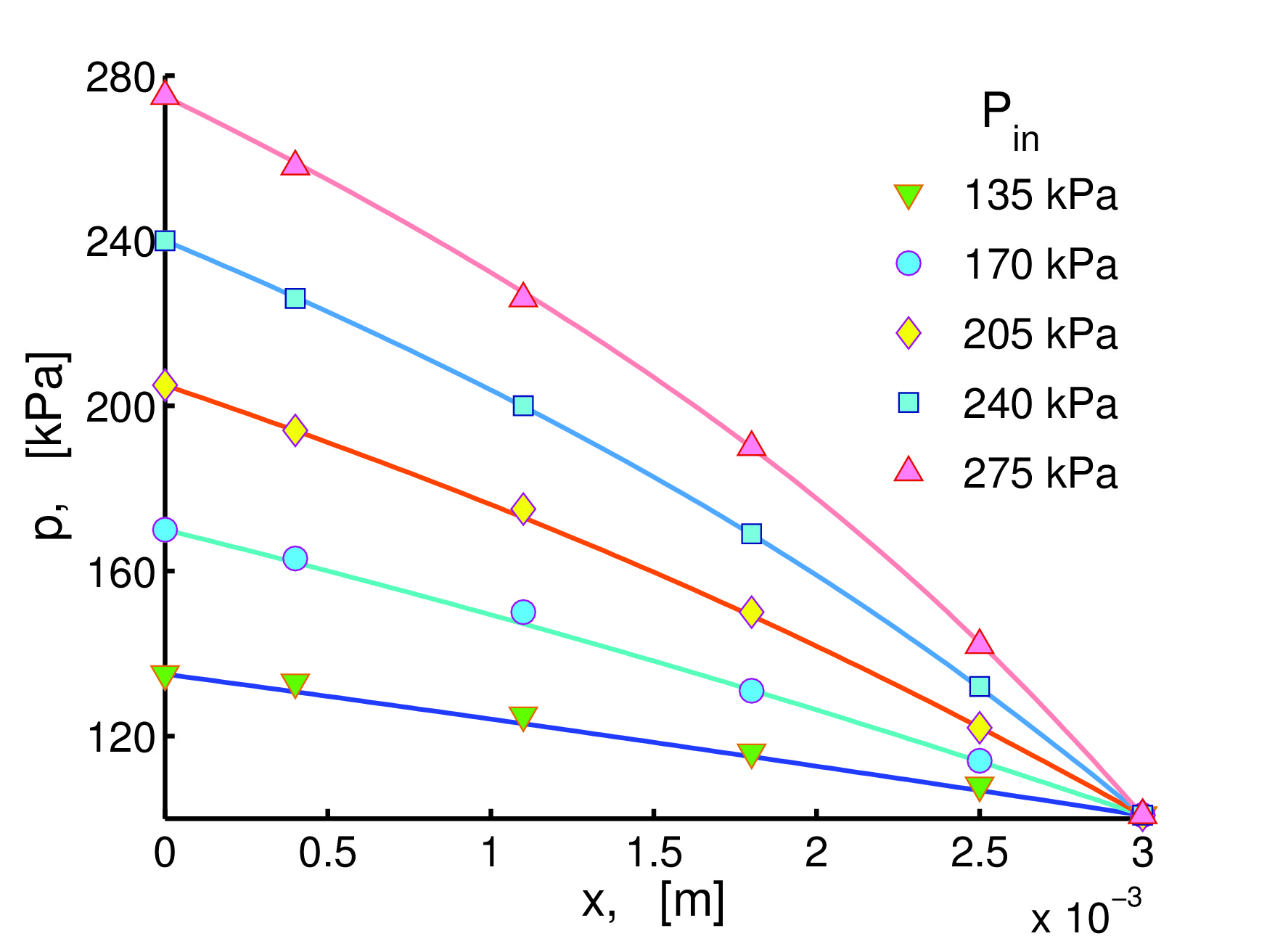}
      \captionsetup{labelformat=empty}
	\caption{Model 12} 	\label{Case2Plot12}
\end{minipage}
      \captionsetup{labelformat=empty}
	\caption{Figure 6(c). (See the caption to Figure 6(a)).}
\end{figure}


\begin{figure}[ht]   
\begin{minipage}[b]{0.6\linewidth}
\includegraphics[width=0.8\linewidth]{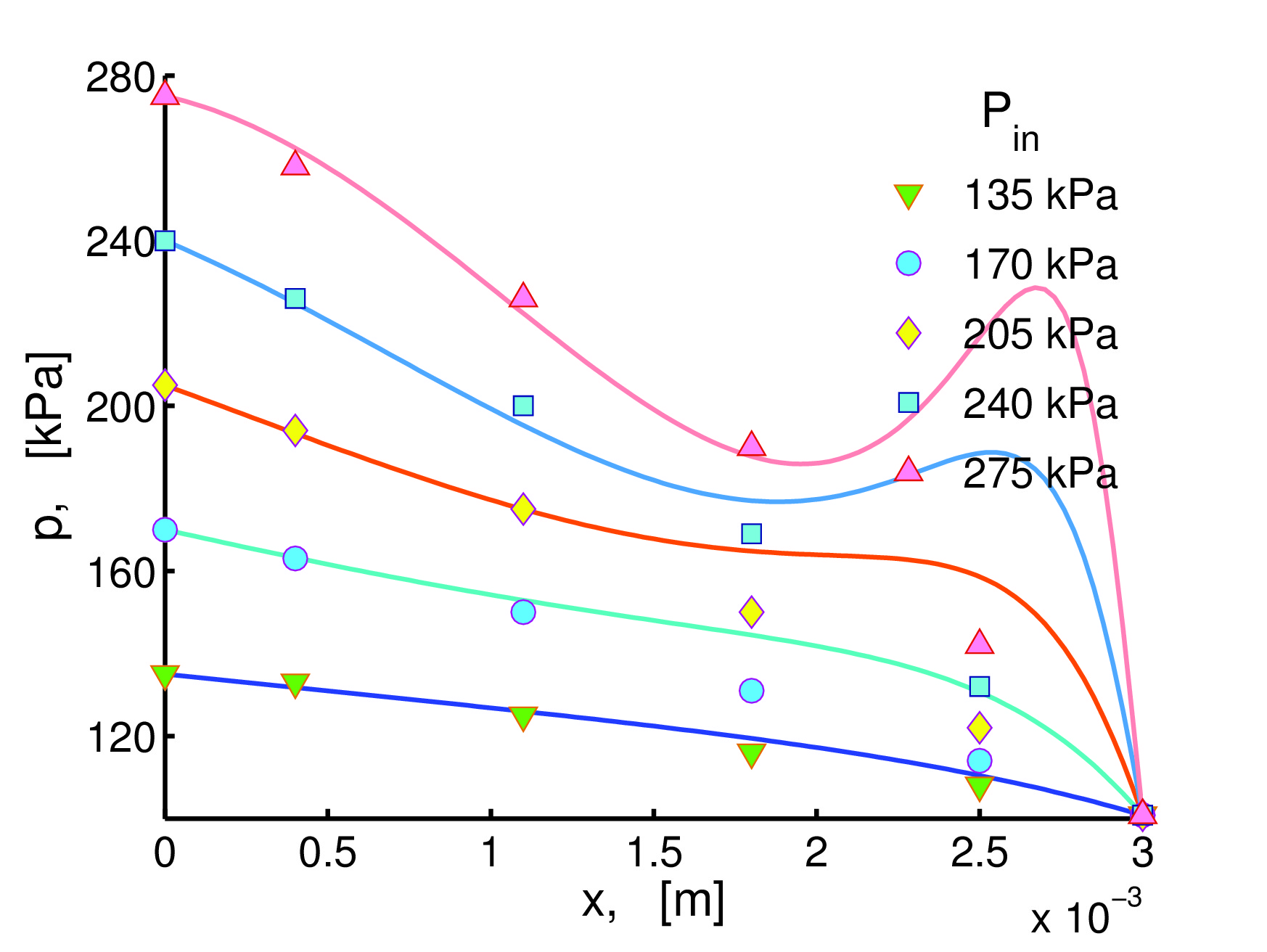}
      \captionsetup{labelformat=empty}
	\caption{Model 13} 	\label{Case1Plot13}
\end{minipage}
\hspace{-1.5cm}
\begin{minipage}[b]{0.6\linewidth}
\includegraphics[width=0.8\linewidth]{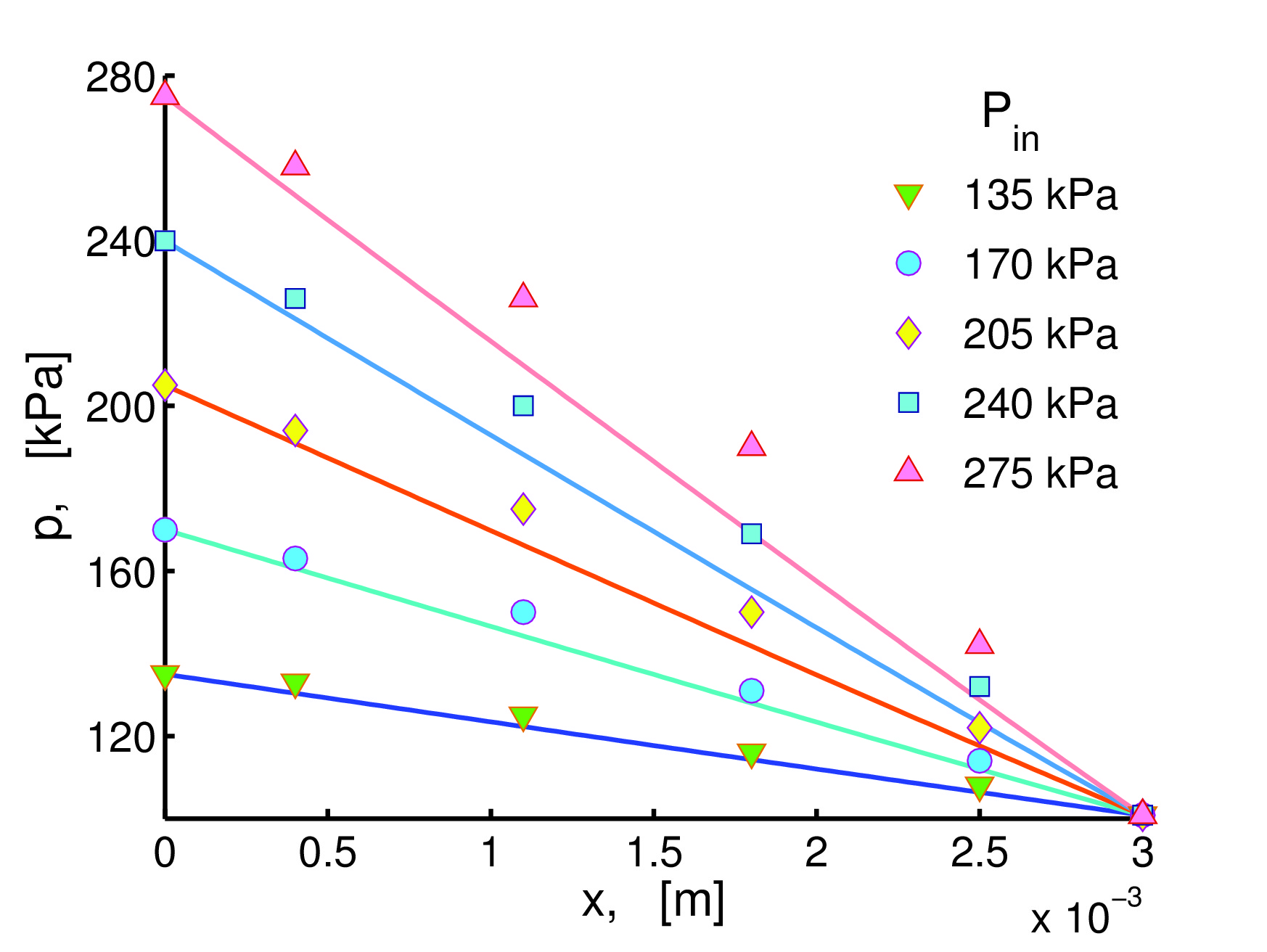}
      \captionsetup{labelformat=empty}
	\caption{Model 14}  	\label{Case2Plot14}
\end{minipage}
\begin{minipage}[b]{0.6\linewidth}
\includegraphics[width=0.8\linewidth]{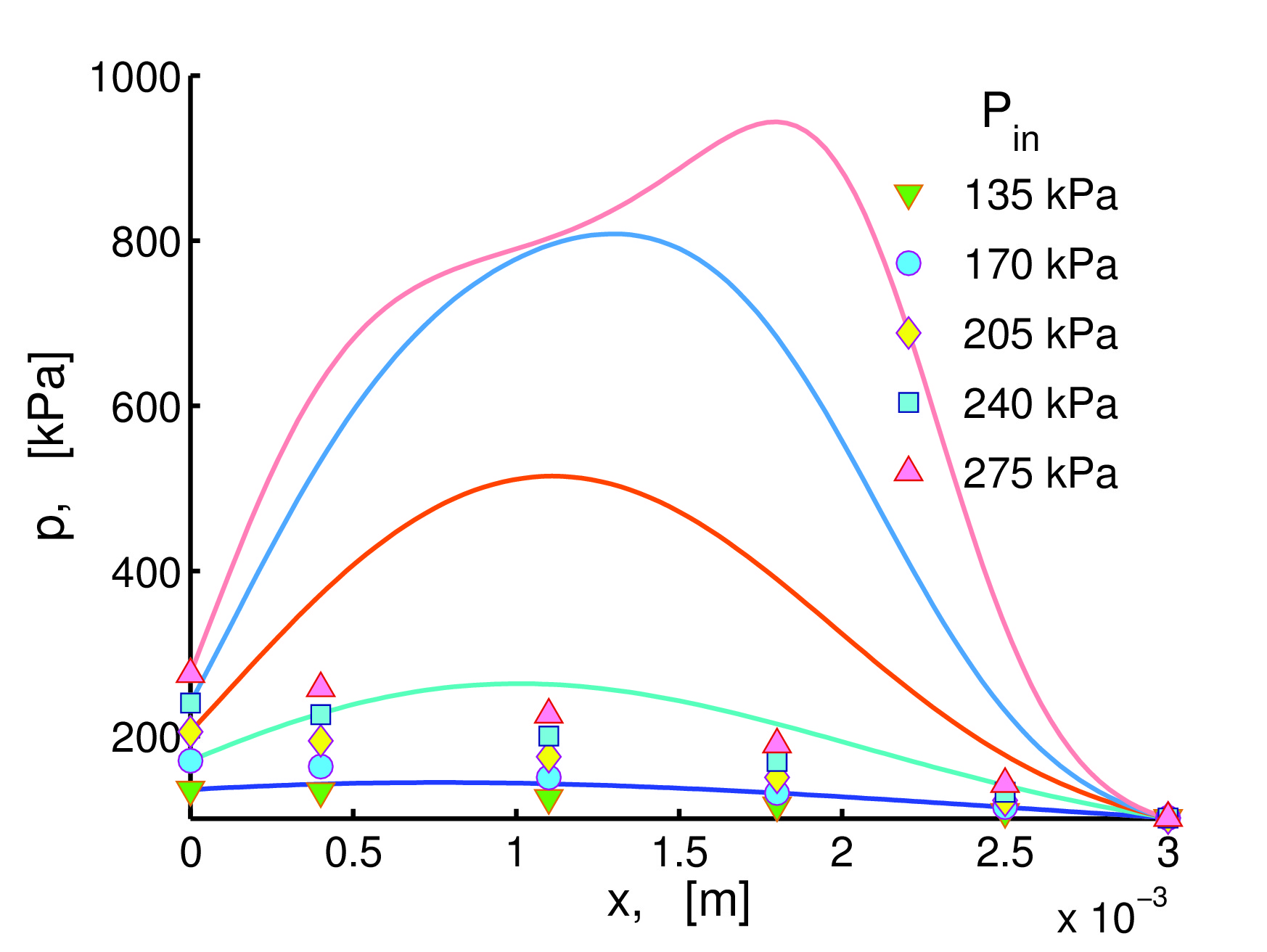}
      \captionsetup{labelformat=empty}
	\caption{Model 15} 	\label{Case1Plot15}
\end{minipage}
\hspace{-1.5cm}
\begin{minipage}[b]{0.6\linewidth}
\includegraphics[width=0.8\linewidth]{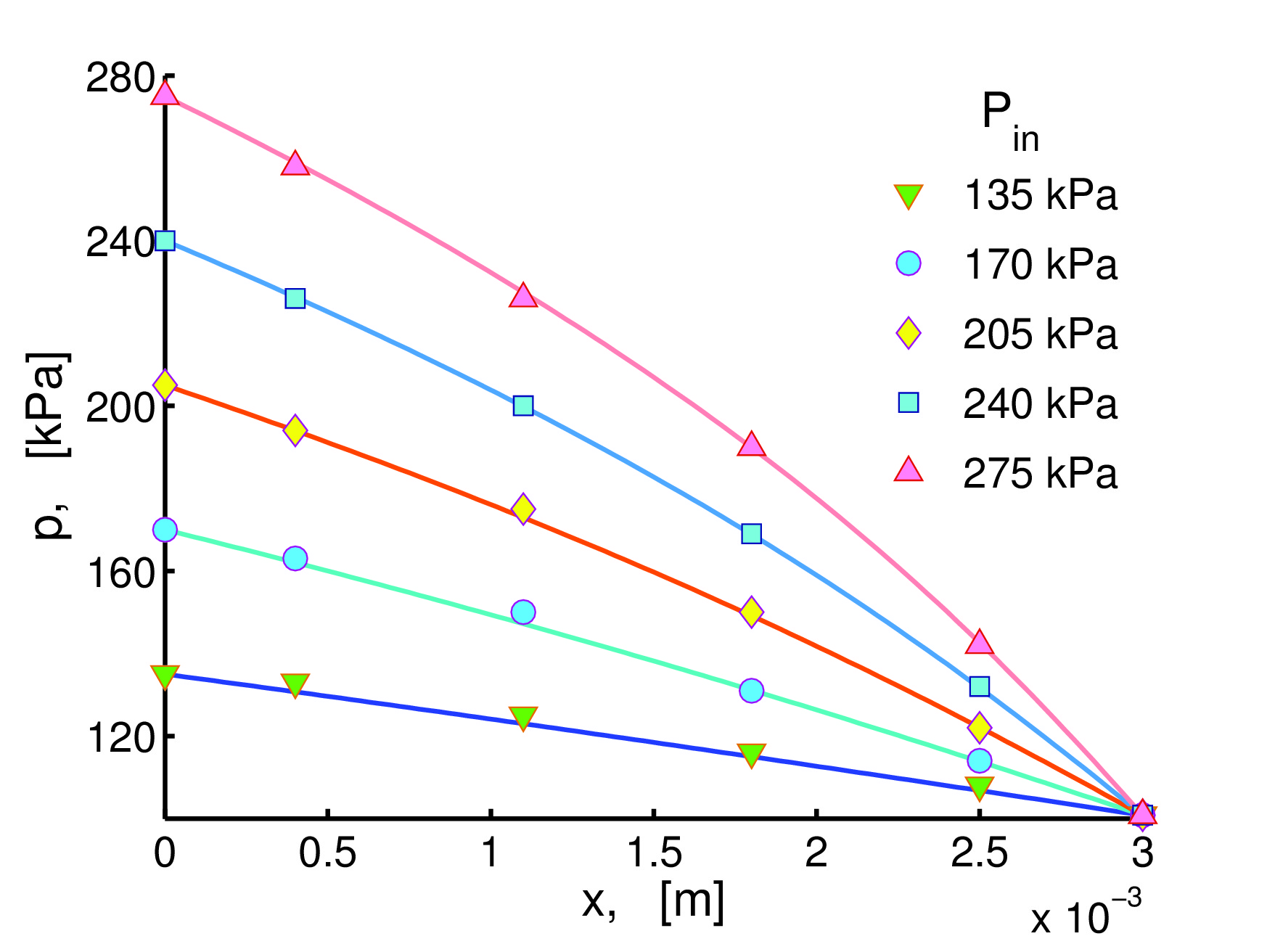}
      \captionsetup{labelformat=empty}
	\caption{Model 16}	\label{Case2Plot16}
\end{minipage}
      \captionsetup{labelformat=empty}
	\caption{Figure 6(d). (See the caption to Figure 6(a)).}
\end{figure}


Only Model 12 and Model 16 show a good match between the numerical solutions and the
experimental data. Some of the models show a fairly good match,  e.g. Model 3 and Model 10;
some of the models show significant errors, e.g. Models 1, 5, 7, 8, and 14; while yet other
models are very badly in error, e.g. Models 2, 4, 6, 9, 11, 13, and 15.

Darcy's law,  Model 1, where all model parameters are taken to be independent of the pressure, gives
linear profiles and is clearly unsatisfactory.

Fig. 7 show the relative error on log-scale for the sixteen models considered.

Model 16 is the case where all the model parameters are pressure dependent.
The error calculated in Model 16 is the smallest  among all the sixteen models.

From these results, we conclude that Model 16 is the best fit among all the models considered; this
demonstrates the importance of retaining all model parameters to be pressure dependent throughout
the simulations in order to obtain the best results.

An exception is for the smallest inlet pressures $P_{in}$, where most of the models yield fair
agreement with the data. This means that low $P_{in}$ should be avoided in such experiments as
the models are not critically sensitive to pressure dependency at low inlet pressures, so estimates
of rocks properties cannot be made with accuracy.

\begin{figure}
\centering
\includegraphics[width=8cm]{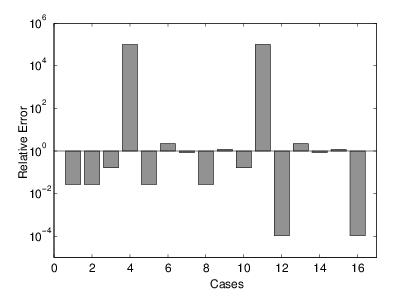}  
      \captionsetup{labelformat=empty}
      \caption{Figure 7. The error, $\epsilon^2$ using equation \eqref{Error}, between the simulated and
                    experimental data, for the 16 models listed in Table \ref{Table02}.}
\label{fig:Error1B}
\end{figure}

From Model 16 we can make estimates of the rock properties. \citet{civan2011shale} took
most of the model parameters to be constant and assumed a constant value for the porosity,
$\phi = 0.2$, independent of pressure, see Table \ref{Table03}. On this basis his model predicted  
a value for the rock permeability to be $K = 10^{-15}$ m$^2$,  (or $10^{6}$ nD).

In the new model  simulations, the porosity is a variable and dependent upon the pressure. From the
Model 16 simulations, it lies in the range $0.1901 \leq \phi \leq 0.2003$, and  it was found that
the permeability lies in the range $10^{-14} \leq K \leq 10^{-15}$ m$^2$, (or $10^6\le K\le 10^7$ nD).
Although these estimates are comparable to Civan's estimate above, these values are much higher
than expected, and not realistic of typical shale rocks.


\subsection{Simulation results with Forchheimer's correction, $B\not=0$}\label{nssm}


We now  include a non-zero turbulence correction factor $B\neq  0$ ($F \neq 1$), equation
\eqref{SSM2a}, to produce a new set of transport models -- we will refer to them as
Models 1 to 16 with $B\not=0$. As the critical importance of retaining the pressure-dependence of all model
parameters has already been established in the previous subsection, here we consider only Model 16,
where all parameters are pressure-dependent. With $B\neq 0$, we have four additional parameters,
$a_{B}$, $b_{B}$, $c_{B}$, and $d_{B}$, that arise in the model from the consideration of the
Forchheimer's correction term.

Fig. 8 shows the simulation results when we take the same parameter values as for Model 16 with
$B=0$, Table \ref{Table04}. The four additional parameters are, initially, guesstimated. The simulations
are significantly in error of the data, except as usual for the lowest $P_{in}=275KPa$.


\begin{table}[ph]
\centering
\addtolength{\tabcolsep}{-2pt} 
\begin{tabular}{p{4cm} p{3cm} }
\multicolumn{2}{c}{New Steady State Model with turbulence correction}  \\
Reservoir  Parameters               & values                           \\
$L$ (m)                             & {0.003}                          \\
$N_x$                               & {100}                            \\
$R_g$ (JKMol/K)                     & {8314.4}                         \\
$M_g$ (Kg/KMol/K)                   & {28.013}                         \\
$T$ (K)                             & {314}                            \\
$p_c$ (KPa)                         & {3396}                           \\
$t_c$ (K)                           & {126.19}                         \\
$b_{SF}$                            & {-1}                             \\
$\sigma_0$                          & {1.3580}                         \\
$A_{\sigma}$                        & {0.1780}                         \\
$B_{\sigma}$                        & {0.4348}                         \\
$a_{\tau}$                          & {1.5}                            \\
$a_{\phi}$                          & {0.10}                           \\
$b_{\phi}$                          & {-0.939e-1}                      \\
$c_{\phi}$                          & {0.39}                           \\
$\alpha_{KC}$                       & {1.0}                            \\
$\beta_{KC}$                        & {0.9}                            \\
$\Gamma_{KC}$                       & {0.72e-8}                        \\
$a_{\beta}$                         & {3.1e01}                         \\
$b_{\beta}$                         & {0.5}                            \\
$c_{\beta}$                         & {1.35}                           \\
$d_{\beta}$                         & {0.4}                            \\
\text{Error}                        & {5.972e-5}                       \\
\end{tabular}	
\vskip 0.2cm	
\caption[Test Data Set]
{Model parameters used in the New Steady State Model.}
\label{Table05}
\end{table}



\begin{figure}
\centering
\includegraphics[width=8cm]{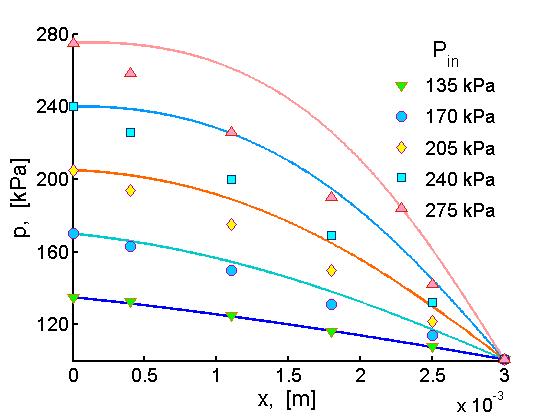}
      \captionsetup{labelformat=empty}
      \caption{Figure 8. Pressure against distance for different inlet pressures, from Model 16
           with $B\not=0$, and with parameter values listed in Table \ref{Table04} .}
\label{Fig8}
\end{figure}

\begin{figure}
\centering
\includegraphics[width=8cm]{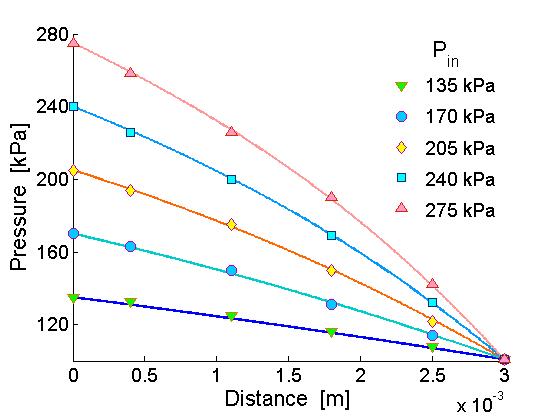}
      \captionsetup{labelformat=empty}
      \caption{Figure 9. Pressure against distance for different inlet pressures, from Model 16
           with $B\not=0$, and with parameter values listed in Table \ref{Table05} . }
\label{Fig9}
\end{figure}


The model parameters were therefore adjusted for the best fit, and the new list of parameter values
is shown in Table  \ref{Table05}. Figure 9 shows the simulation results for different inlet
pressures. We observe an excellent match between the numerical solutions and the experimental data.
The the relative error between the simulated and the measured pressure values is $5.72\times 10^{-5}$,
which is smaller than from Model 16 with $B=0$, in the previous subsection 6.5, Fig. 6(d).

Importantly, the range of porosity was found to lie in the range $0.10<\phi<0.1038$, and the
intrinsic permeability was found to lie in the range $106<K<111$ nD. These are much more realistic
of shale rocks than any previous model. 

\section{Discussion and Conclusions}\label{dac}

In this work, a fully pressure-dependent nonlinear transport model for the flow of shale gas in
tight porous media has been derived accounting for the important physical processes that
exist in the system, such as continuous flow, transition flow, slip flow, surface diffusion, adsorption
and desorption in to the rock material, and also including a nonlinear correction term for high flow
rates (turbulence). This produces an advection-diffusion type of partial differential equation (PDE) with
pressure dependent model parameters and associated compressibility coefficients, and with highly
nonlinear apparent convective velocity and apparent diffusivity. The model was developed initially
for the general case of transient flow of single phase gas flow in a three-dimensional porous system
with gravity and a general source term.

A steady state one-dimensional version of the model without gravity and without external source was
used to  determine the rock properties by matching the pressure distrbution across a shale rock core
sample  obtained from pressure-pulse decay tests for different inflow pressure conditions. It was found
that when the high flow rate correction factor is also excluded ($B=0$), then the model with fully pressure
dependent parameters, Model 16, still gives the least errors compared the data, and yields estimates
of rock properties that are in close agreement with Civan's model; however, both estimates are
not realistic of typical shale rocks.

When the high flow rate correction factor is included ($B\not=0$) in the model, the errors are
further reduced, and the estimates for the porosity and permeability are much improved and are 
within the known range of shale rock properties, we believe for the first time. The estimates are 
more realistic than obtained from previous transport model. This is a noteable achievement for the 
present model development and application, and sets a benchmark for future development.

We can draw the following conclusions. Firstly, a realistic transport model should incorporate all of the
important physical transport sub-processes in  the porous system. Secondly, model parameters and
associated compressibility coefficients should be pressure dependent throughout the numerical 
procedure. Thirdly, the a Forchchiemer correction term for high flow rates is very important for good 
estimation of rock properties. The simulation results presented here for estimating  rock properties  
illustrate the potential of the modelling startegy presented here in producing accurate simulations 
of gas flow in shale gas reservoirs.

In the future, the second phase of this work is to incorporate this model in to a new model 
for gas transport through fractured media, such as tight shale rocks.

\section*{Acknowledgements}
The authors would like to acknowledge the support provided by King Abdulaziz City for Science and Technology (KACST) through the National Science, Technology and Innovation Plan (NSTIP), and
through the Science Technology Unit at King Fahd University of Petroleum \& Minerals (KFUPM) for funding this work through project No. 14-OIL280-04.

\bibliographystyle{model2-names}

\end{document}